\documentclass[
  journal=pasa,
  manuscript=article-type,
  year=2024,
]{cup-journal}

\usepackage[nopatch]{microtype}
\usepackage{booktabs}
\usepackage{amsmath}
\usepackage{natbib}
\usepackage{aas-macros}
\usepackage[utf8]{inputenc} 
\usepackage[T1]{fontenc}    
\usepackage{hyperref}       
\usepackage{url}            
\usepackage{booktabs}       
\usepackage{amsfonts}       
\usepackage{amssymb}
\usepackage{rotating}
\usepackage{nicefrac}       
\usepackage{microtype}      
\usepackage{lipsum}
\usepackage{graphicx}       
\graphicspath{{media/}}     
\usepackage{tabularx}
\newcommand{\minus}{\scalebox{0.75}[1.0]{$-$}}

\title{$\chi^2$ from Redundant Calibration as a Tool in the Detection of Faint Radio-frequency Interference}

\author{Theodora Kunicki}
\affiliation{Department of Physics, Brown University, Providence, RI, 02912, USA}
\email[Theodora Kunicki]{theodora\_kunicki@brown.edu}

\author{Jonathan C. Pober}
\affiliation{Department of Physics, Brown University, Providence, RI, 02912, USA}


\begin{document}

\begin{abstract}
    Radio-frequency interference detection and flagging is one of the most difficult and urgent problems in 21 cm Epoch of Reionization research.
    In this work, we present $\chi^2$ from redundant calibration as a novel method for RFI detection and flagging, demonstrating it to be complementary to current state-of-the-art flagging algorithms.
    Beginning with a brief overview of redundant calibration and the meaning of the $\chi^2$ metric, we demonstrate a two-step RFI flagging algorithm which uses the values of this metric to detect faint RFI.
    We find that roughly 27.4\% of observations have RFI from digital television channel 7 detected by at least one algorithm of the three tested: 18.0\% of observations are flagged by the novel $\chi^2$ algorithm, 16.5\% are flagged by SSINS, and 6.8\% are flagged by AOFlagger (there is significant overlap in these percentages).
    Of the 27.4\% of observations with detected DTV channel 7 RFI, 37.1\% (10.2\% of the total observations) are detected by $\chi^2$ alone, and not by either SSINS or AOFlagger, demonstrating a significant population of as-yet undetected RFI.
    We find that $\chi^2$ is able to detect RFI events which remain undetectable to SSINS and AOFlagger, especially in the domain of long-duration, weak RFI from digital television.
    We also discuss the shortcomings of this approach, and discuss examples of RFI which seems undetectable using $\chi^2$ while being successfully flagged by SSINS and/or AOFlagger.
\end{abstract}


\section{Introduction}\label{introduction}
During the Universe's Epoch of Reionization (EoR), which took place when $20 \gtrsim z \gtrsim 6$, the intergalactic medium (IGM) transitioned from being almost entirely neutral to an effectively completely ionized state.
While the process of reionization is generally expected to have been driven by ionizing radiation from early stars, this period of the Universe's history has yet to be directly observed.
The details of reionization are sensitive to other important cosmological factors such as the process of early galaxy formation, so its observation is of high potential value to experimental cosmology.
Two review articles which describe the Epoch of Reionization in depth are \cite{choudhury_2022} and \cite{wise_2019}.

The most promising experimental probe into the EoR is 21 centimeter radiation emitted by neutral hydrogen gas in the IGM. 
Neutral hydrogen gas emits radiation at a wavelength of 21 centimeters owing to the spin-flip transition in the atom; notably, this transition requires both an electron and a proton, so 21 cm radiation is \emph{not} emitted by ionized hydrogen.
By mapping the EoR's 21 cm radiation, cosmologists hope to learn about the spatial distribution of neutral hydrogen in the IGM as the EoR progressed.
For reviews of 21 cm cosmology, see, for example, \cite{morales_2010}, \cite{furlanetto_2006}, \cite{liu_2020} and \cite{pritchard_2012}.

Efforts to observe the 21 cm power spectrum from the EoR are underway at LOFAR\footnote{\url{https://www.astron.nl/telescopes/lofar/}} \citep{lofar}, HERA\footnote{\url{https://reionization.org/}} \citep{hera}, and the Murchison Widefield Array\footnote{\url{https://www.mwatelescope.org/}} \citep{tingay_2013,mwa_phase_ii}.

Detecting the 21 cm EoR power spectrum is difficult for multiple reasons, and chief among them is the contamination introduced to the data by anthropogenic radio-frequency interference (RFI).
Removing RFI-contaminated data from a power-spectrum analysis is essential if the measurement is to be sensitive to the 21 cm EoR signal \citep{wilensky_2020}.
While LOFAR is situated in the Netherlands, where RFI is omnipresent, HERA and the MWA are located in more radio-quiet sites in South Africa and Western Australia, respectively.
Even in these radio-quiet sites, there are many sources of RFI, including digital television (DTV) (which may even be reflected by passing aircraft \citep{absolving_ssins}), ORBCOMM satellite transmissions \citep{sokolowski}, and FM radio, which may reflect from satellites \citep{zhang}.
The MWA, in particular, is located far from any transmitters, and still observes DTV regularly, as well as narrow- and broad-band RFI from time to time \citep{offringa_2015}.

There are two software packages regularly used in RFI detection at the MWA, AOFlagger \citep{offringa_2015}, and SSINS \citep{absolving_ssins}.
While these software tools are both powerful, it has been demonstrated that they fail to excise all RFI, leaving behind so-called ``ultra-faint'' RFI in the data \citep{wilensky_23}.
In \cite{wilensky_23}, statistical methods were used to demonstrate that this ``ultra-faint'' RFI is present in existing datasets that have been used for analysis, and that even this very weak residual RFI is extremely deleterious to the effort to accurately measure the 21 cm EoR power spectrum.

In a radio interferometer, a baseline is defined as the displacement vector which separates two antennas whose signals are to be correlated by the instrument.
Both HERA and the MWA have baseline redundancies, i.e., there are baseline vectors which are repeated multiple times throughout the array.
This baseline redundancy is visible as translationally symmetric layouts in all (in the case of HERA) or a subset (in the case of the MWA) of antennas\footnote{Antennas in the MWA are usually referred to as ``tiles'' and consist of 16 crossed dipoles of less than a meter in any dimension arranged in a $4\times4$ grid over a ground screen. In this work, we will use the terms ``tile'' and ``antenna'' interchangeably to refer to MWA tiles.}.
Baseline redundancy was built into these interferometers in order to increase sensitivity on these baselines \citep{parsons_et_al_2012a}, and also to make use of redundant calibration \citep{wieringa_1992}, which uses the fact that identical baseline vectors should measure identical visibilities. 
In a redundant calibration algorithm, a metric called $\chi^2$, which measures a baseline's visibility's noise-normalized square deviation from a calculated ``consensus'' visibility for that baseline group, is minimized for each baseline by adjusting antenna gains (see Section \ref{redundant} for a formal definition of the $\chi^2$ used in redundant calibration). 

In this work we present an alternative use of the $\chi^2$ metric from redundant calibration as an RFI detection tool.
This approach has been proposed before in \cite{MITEoR}, but has not yet been investigated as a standalone technique.
\cite{li_2019} observed elevated $\chi^2$ in MWA data corresponding to RFI, and used it to hand-flag some contaminated data, but did not include a systematic search using $\chi^2$ to automatically flag data or a comparison to the RFI detected by SSINS and AOFlagger.
In this work, we will demonstrate that $\chi^2$ from redundant calibration is not only effective at detecting RFI in MWA data, but can also detect RFI which is not visible to either SSINS or AOFlagger.

The structure of this paper is as follows: Section \ref{data_and_calibration} describes our dataset and preprocessing, provides an overview of redundant calibration, and introduces the $\chi^2$ metric, which will be the focus of this work.
Section \ref{RFI_identification} discusses $\chi^2$'s sensitivity to RFI, and describes our method used to generate flags based on RFI detected in $\chi^2$.
Section \ref{other_algorithms} briefly describes AOFlagger and SSINS, the performance of which is to be compared to $\chi^2$ flagging.
Section \ref{results} presents the results of our analysis, including describing the specific RFI that is detected with this method, and presenting a comparison of $\chi^2$ with AOFlagger and SSINS on their performance flagging the most common types of RFI observed in the dataset.
Section \ref{discussion_and_conclusion} provides a discussion of the results and their implications for 21 cm cosmology, and concludes the work.
This paper has three short appendices; in \ref{effects_of_time_averaging}, we discuss the effects of time-averaging on $\chi^2$ generated by redundant calibration, in \ref{redcal_run}, we discuss the specific parameters used to run redundant calibration, and \ref{appendix} is a collection of interesting and exemplary RFI events, visualized using all three flagging algorithms.

\section{Data and Redundant Calibration}\label{data_and_calibration}


\subsection{Data}\label{data}
Data were downloaded from the MWA All-Sky Virtual Observatory (ASVO)\footnote{\url{https://asvo.mwatelescope.org/}} service in MWA correlator FITS format.
The data lie in the band between 167.055 MHz and 197.735 MHz, and were collected by the MWA between the dates of October 15\textsuperscript{th} and December 15\textsuperscript{th} of 2016.
This period coincides with the beginning of MWA Phase II operations, and these data were taken with the array in its 128-antenna ``compact'' configuration \citep{mwa_phase_ii}, which includes 71 antennas in two redundant, compact hexagons of 36 and 35 antennas (both hexagons were intended to comprise 36 antennas, but one antenna location was inaccessible on site \citep{li_18}).

\begin{figure}[ht!]
\caption{Antenna positions in the MWA Phase II compact configuration.}
\label{fig:antpos}
\centering
\includegraphics[width=\textwidth]{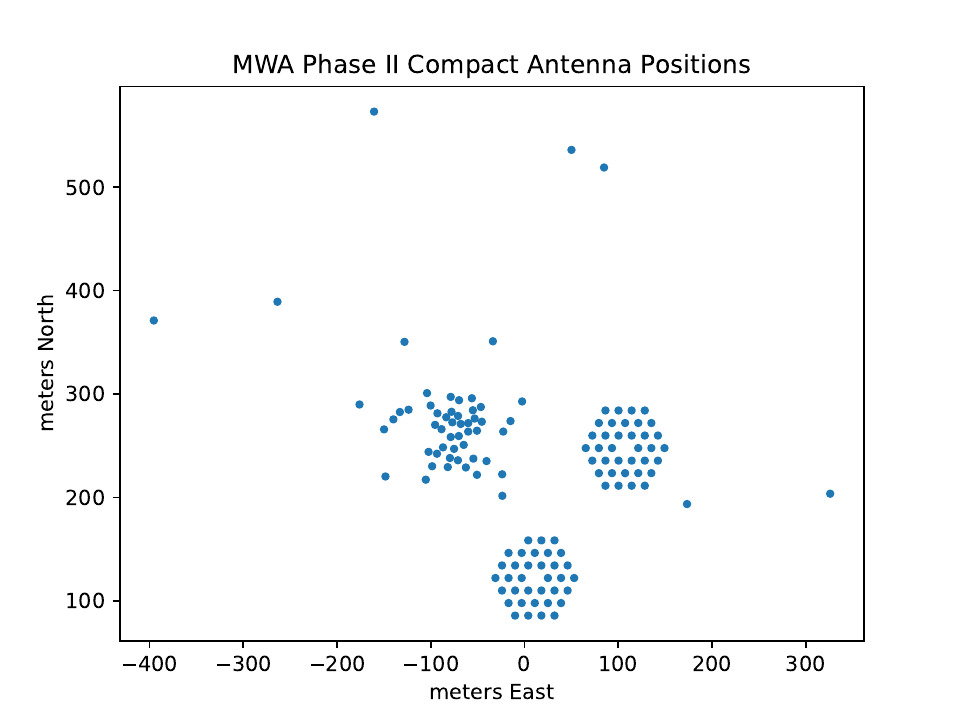}
\end{figure}

Figure \ref{fig:antpos} illustrates the antenna positions of the Phase II compact configuration of the MWA.

Within the dataset, there are observations of both the EoR0 (R.A. = 0.0\textdegree, Dec. = -27.0\textdegree) and EoR1 (R.A. = 60.0\textdegree, Dec = -30.0\textdegree) fields \citep{jacobs_2016}.

The EoR0 observations in this set are the same data analyzed in \cite{li_2019}, and (apart from three nights of data analyzed in \cite{zhang_2020}) the EoR1 data are heretofore unanalyzed.
The dataset comprises 51.8 hours of observations, spanning 1665 approximately two-minute observation files.

For both the EoR0 and EoR1 datasets, the data include several pointings of the MWA, which are accomplished instrumentally by introducing relative delays into the analog signals from each of the 16 individual dipoles which make up each MWA tile, changing the angle of the beam's phase center on the sky.
In a single night of observing, the same field on the sky is tracked by allowing it to drift through the beam several times as the beam is phased in discrete increments to follow it.
According to the grid numbering scheme in \cite{beardsley_2016}, the pointings included in the dataset are $\minus 3$ to $+2$ for the EoR0 observations, and $\minus 3$ to $+3$ in the EoR1 observations (where pointing 0 is an observation with the beam center at zenith).

The MWA correlator used during Phase II of operations introduced a coarse-band structure to the data, which is discussed in detail in Section \ref{flagging}.
Briefly, due to the design of the correlator, there are certain frequency bins (corresponding to the edges and centers of 24 coarse bands) which always contain corrupted data.
Throughout this paper, we will mask out corrupted data in these bins, and will always note when an algorithm has been modified to work around these masked sections of data.


\subsection{Data Preprocessing and Redundant Calibration}\label{preprocessing}
Before they are run through redundant calibration, the data are downsampled in time using the python package pyuvdata\footnote{Available at \url{https://github.com/RadioAstronomySoftwareGroup/pyuvdata}} version 2.4.0 \citep{pyuvdata}.
The files are time-averaged from a time cadence of one visibility every 2 seconds to one visibility every 18 seconds; this is to reduce noise in the observations, which at a 2-second cadence prove too noisy to allow the redundant calibration algorithm to arrive at reasonable gain solutions.  The rationale for this time-averaging is discussed further and example images are provided in \ref{effects_of_time_averaging}.

Pyuvdata is likewise used to select down to only the 71 antennas in the redundant section of the array.
This is because the software package used in this work for running redundant calibration, \verb|hera_cal|, was developed for use with the HERA telescope, which is fully redundantly laid out \citep{dillon_and_parsons_2016}, as opposed to the MWA which has some redundant antennas alongside pseudorandomly positioned antennas.
The software is therefore likely to calculate spuriously ``redundant'' baselines when all of the pseudorandomly positioned antennas are included.

\subsection{Redundant calibration}\label{redundant}
Once the data have been preprocessed, \verb|hera_cal| \footnote{Available at \url{https://github.com/HERA-Team/hera_cal}, version 3.4.1.dev2+g102e14f8 was used in this work.} \citep{Dillon_2020} is used to run redundant calibration, using \verb|hera_cal|'s \verb|redcal_run()| method.


Redundant calibration takes advantage of the fact that baselines which have the same vector separating their component antennas should in principle measure the same visibility from the sky.
We assume that a measured visibility for a given time $t$, frequency $\nu$, and polarization $p$, between antennas $i$ and $j$, $v^{}_{ij}$ can be expressed as the product of two antenna-dependent complex gains, $g^{}_i$ and $g^{}_j$ and a ``true'' sky visibility $y^{}_{ij}$, with an added noise term $n^{}_{ij}$
\begin{align} \label{calibration_eq}
v^{}_{ij}(t, \nu, p) \approx g^{}_{i}(t, \nu, p)g^*_{j}(t, \nu, p)y^{}_{ij}(t, \nu, p) + n^{}_{ij}
\end{align}
We drop the explicit time, frequency, and polarization dependencies in much of the following discussion, but it should be emphasized that redundant calibration provides an independent solution for the gains and visibilities at every unique combination of time, frequency, and polarization.  This also means that we get a value for $\chi^2$ at each time, frequency, and polarization.

The goal of calibration is to solve for the gains $g^{}_i$.
We go about this by minimizing $\chi^2$, defined as
\begin{align} \label{chisq}
\chi^2 &= \sum_{i<j}\frac{\left|v^{}_{ij} - g^{}_ig^*_jy^{}_{ij}\right|^2}{\sigma^2_{ij}}
\end{align}
Where $y^{}_{ij}$ has subtly changed in meaning to be a ``consensus'' visibility calculated for the baseline group (a ``baseline group'' being a collection of baselines with identical baseline vectors), and $\sigma^2_{ij}$ is an estimation of the noise variance of that baseline type.
The consensus visibility, $y^{}_{ij}$, of a group of identical baselines is a free parameter to be solved for in our system of linear equations, alongside the gains and visibilities, and its value is such that the overall $\chi^2$ values for that baseline group are minimized.

Minimizing these $\chi^2$ values is our objective as we adjust parameters in a large system of linear equations. In the redundant subsection of the MWA, there are 71 antennas, and if we only consider redundant baseline groups containing two or more baselines, that leaves 181 unique baseline types; 71 complex gains and 181 complex visibilities make for 252 complex free parameters to solve for, but the number of equations is much larger at $(71 \times 70)/ 2 = 2485$ (the total number of baselines in the redundant subsection of the array), making this system overdetermined.

$\chi^2$ and its sensitivity to radio-frequency interference will be the focus of this work.
In order to obtain $\chi^2$ for analysis, we must run redundant calibration using \verb|hera_cal|. 
However, it is vital to note that the results of this redundant calibration process --- the gains and visibilities generated by the algorithm --- will not be used in our analysis.
Only the $\chi^2$ values are of interest in finding and flagging RFI.

There are three different algorithms, run in series on the uncalibrated data, used to solve this system of equations in \verb|hera_cal|.
The first is aptly named \verb|firstcal|, the second is known as \verb|logcal|, and the third is called \verb|omnical|\footnote{Further complicating the history of these algorithms, the software package described in \cite{MITEoR} is called \texttt{OMNICAL}, and contains three algorithms: \texttt{logcal}, \texttt{lincal}, and \texttt{omnical} (where we use lowercase letters to distinguish the algorithm \texttt{omnical} from the software package \texttt{OMNICAL}).  \texttt{lincal} and \texttt{omnical} perform the same task with different solvers.  Previous work applying redundant calibration to the MWA (\cite{li_18}, \cite{li_2019}, and \cite{zhang_2020}) use the \texttt{OMNICAL} package to run \texttt{logcal} and \texttt{lincal} but not \texttt{omnical}.  To be explicit, in this work we use the \texttt{hera\_cal} software package (not \texttt{OMNICAL}) and the \texttt{omnical} algorithm (not \texttt{lincal}).}.
For an in-depth description of how these algorithms work, please consult \cite{li_18} and \cite{Dillon_2020}.

\begin{figure*}[t!]
\caption{The mean value of $\chi^2/\mathrm{nDoF}$ versus local sidereal time; each point represents one two-minute observation, with the standard deviation of $\chi^2/\mathrm{nDoF}$ within the observation represented as an error bar. Each color represents data from a different night, while each shape represents data from a different pointing.  The $\chi^2$ means have a pronounced LST dependence. Within individual instrument pointings, there is also an upward trend. Because of the dependence on LST, a single $\chi^2/\mathrm{nDoF}$ cutoff value applied across all observations in sub-optimal for flagging RFI.}
\label{fig:LST_v_mean}
\centering
\includegraphics[width=\textwidth]{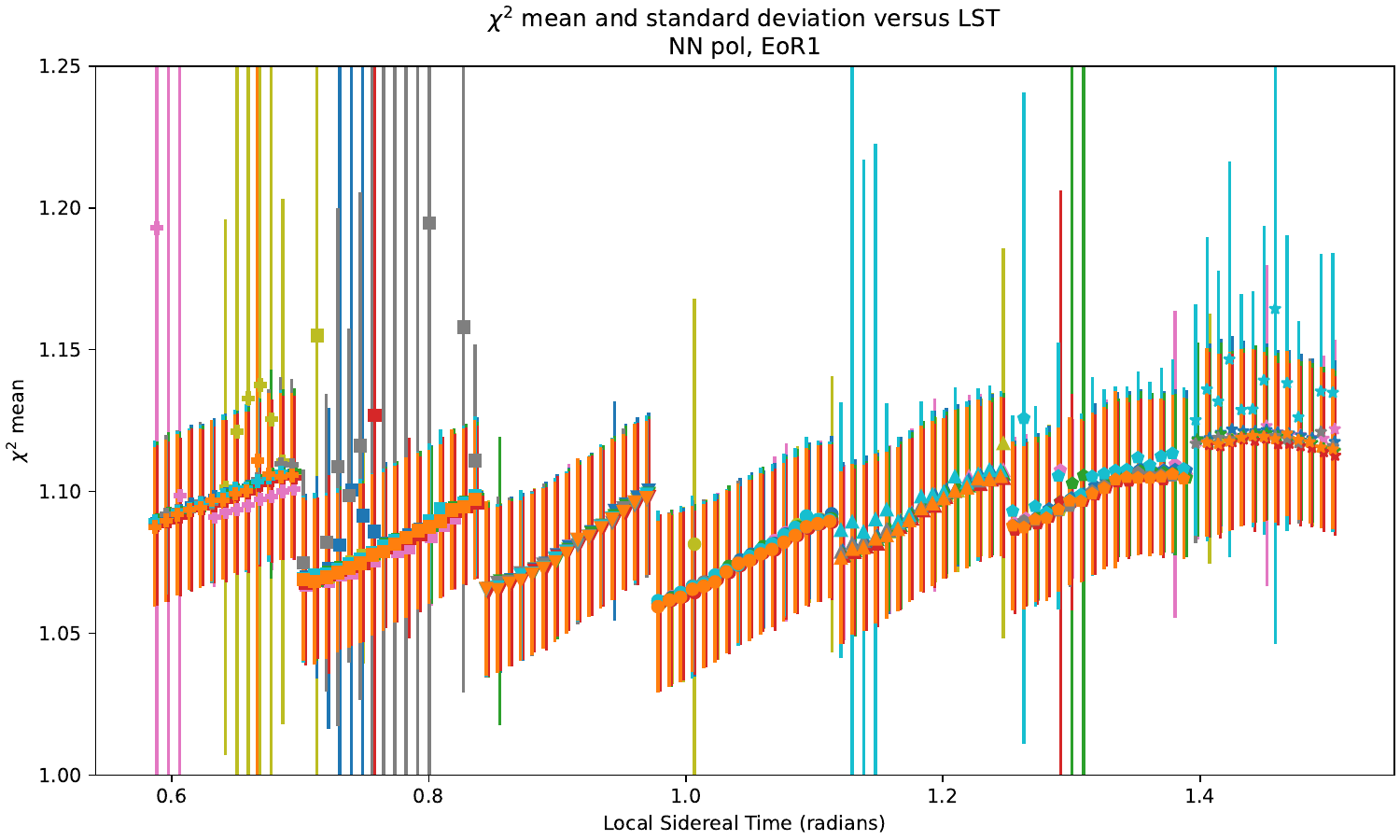}
\end{figure*}

\begin{figure*}[t!]
\caption{This plot represents the same data as Figure \ref{fig:LST_v_mean}, except now modified z-scores have been taken of the $\chi^2/\mathrm{nDoF}$ data. The mean of the modified z-scores of $\chi^2/\mathrm{nDoF}$ is plotted against local sidereal time, with the standard deviation of the modified z-scores of the data represented as error bars. The LST dependence of $\chi^2$ present in Figure \ref{fig:LST_v_mean} is effectively eliminated by using modified z-score.  As with Figure \ref{fig:LST_v_mean}, color denotes the day data were taken, and different point shapes represent different antenna pointings of the instrument.}
\label{fig:LST_v_mean_z-score}
\centering
\includegraphics[width=\textwidth]{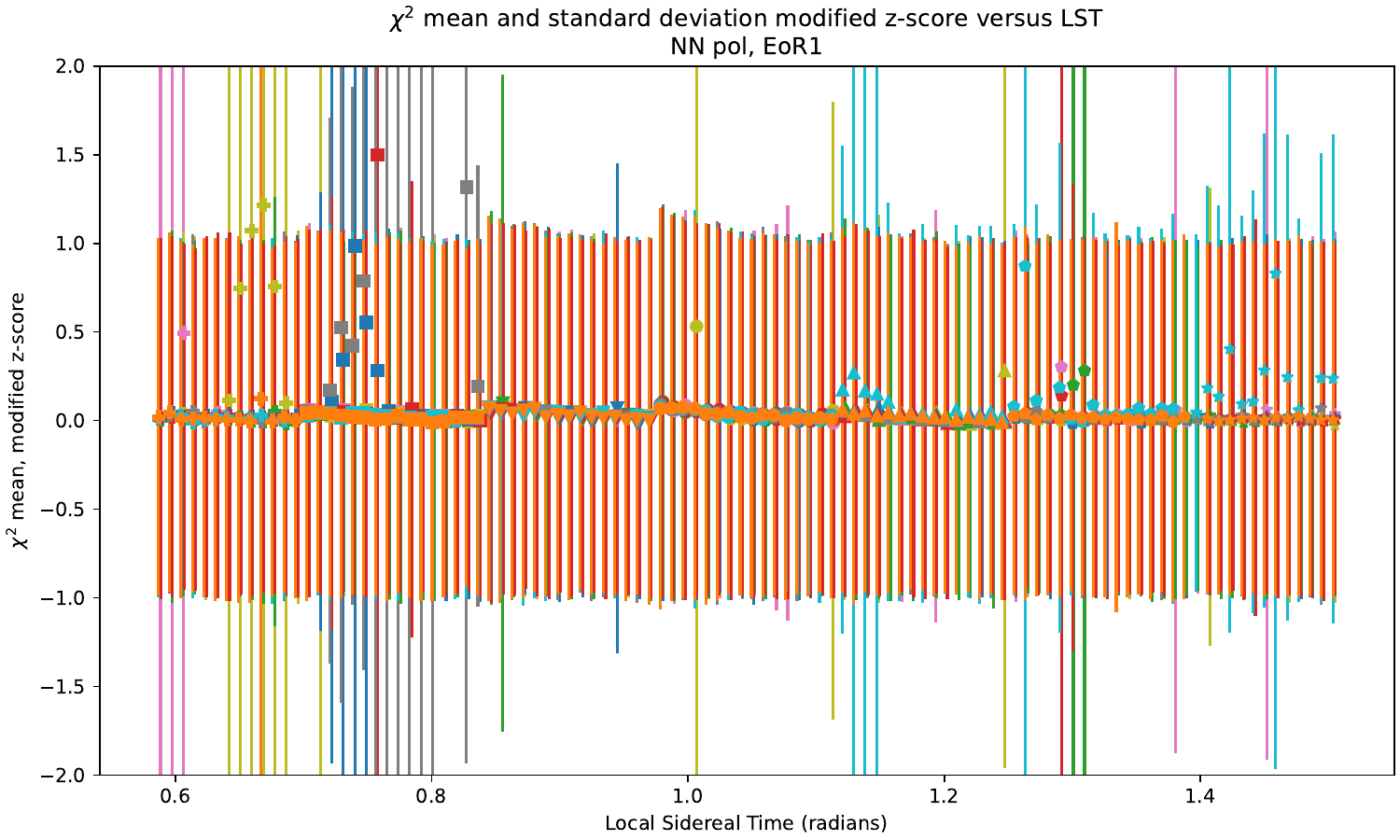}
\end{figure*}
\section{RFI identification and flagging with \texorpdfstring{$\chi^2$}{chi2}}\label{RFI_identification}
\subsection{\texorpdfstring{$\chi^2$}{chi2} and Degrees of Freedom}\label{rfi_characteristics}
As \verb|hera_cal| implements the above algorithms to estimate gains $g^{}_i$ and visibilities $y^{}_{ij}$, it does so with the goal of minimizing the metric $\chi^2$, defined in equation \eqref{chisq}.
This is the overall $\chi^2$ for the entire array (at each time, frequency, and polarization), found by summing over each baseline $ij$, comparing a baseline's measured visibility with the visibility calculated by multiplying the antenna gains $g^{}_i$ and $g_j^*$ by the calculated ``consensus'' visibility $y^{}_{ij}$.
The difference of these quantities is squared, and then divided by an estimation of the noise variance $\sigma^2_{ij}$.
If the measured visibility and calculated visibility only differ by a factor of noise, this quotient should be of order $1$.
When summed over the entire system, we would expect $\chi^2$ of a well-calibrated array to be of order $\mathrm{nDoF}$, or the number of complex degrees of freedom in the system, calculated by \cite{Dillon_2020} to be
\begin{align}\label{dof}
\mathrm{nDoF} &= N^{}_{bl} - N^{}_{ubl} - N^{}_{\mathrm{ants}} + 2.5
\end{align}
Where $N^{}_{bl}$ is the total number of baselines, $N^{}_{ubl}$ is the number of unique baseline types, or alternatively the number of $y^{}_{ij}$ that need to be solved for, and $N^{}_{\mathrm{ants}}$ is the total number of antennas, understood as the number of individual gains $g^{}_i$ to be solved for.

The additive factor of $2.5$ accounts for the degeneracies of redundant calibration in the array, which reduce the total number of parameters which can be solved for in our system of linear equations, increasing the total number of degrees of freedom.
In \cite{Dillon_2020}, this constant factor is calculated to be 2 in a classical redundant array.
In such an array, these 2 complex numbers or 4 real parameters correspond to overall amplitude, overall phase, tip/tilt of the array in the N/S direction, and tip/tilt of the array in the E/W direction.
However, due to having two spatially separated redundant hexagons, the MWA introduces an additional real degenerate parameter --- the phase separation between the two redundant hexagons --- which increases the number of real degeneracies in the system from 4 to 5, and the number of complex degenerate parameters in turn from 2 to 2.5 \citep{Dillon_personal}.

Taking $\chi^2 / \mathrm{nDoF}$, we expect a number close to $1.0$ for each time and frequency if redundant calibration has been successfully run and the resulting calibration is accurate.


$\chi^2$ is intrinsically a measure of how redundant the array actually is.
Said another way, $\chi^2$ measures how well the calibrated baseline groups agree with each other on a measured visibility; in a perfectly calibrated, perfectly constructed array, each baseline should measure identical visibilities.
Higher $\chi^2$ values for a given time or frequency indicate that nominally ``redundant'' baseline groups do not agree on their measured visibilities.

We have observed that $\chi^2 / \mathrm{nDoF}$ is elevated by radio-\\frequency interference (c.f. the lower plot in Figure \ref{fig:noisy_chisq}).
There are two main ways this could occur: either the presence of RFI increases $\chi^2$ by increasing non-redundancy (the numerator of $\chi^2$), or the presence of RFI leads to an underestimation of noise variance $\sigma^2_{ij}$ (the denominator of $\chi^2$).

In order to determine which of these mechanisms was responsible for heightened $\chi^2$ in the presence of RFI, we separated the numerator and denominator of equation \eqref{chisq}.
Within \verb|hera_cal|, the antenna autocorrelations, which measure the total power incident on each antenna element, are used as a proxy for $\sigma^2$. We do find that some powerful RFI events slightly elevate the total power seen in the autocorrelations, and hence elevate \verb|hera_cal|'s estimation of $\sigma^2$. However, this has the effect of suppressing $\chi^2$, opposite to the overall effect in $\chi^2$ we report here.
It follows that RFI elevates $\chi^2$ by magnifying or exacerbating non-redundancy in the array, i.e., it increases the numerator of equation \eqref{chisq}.

\subsection{Modified z-score as a tool to flag outlying \texorpdfstring{$\chi^2$}{chi2} data}

The simplest approach to flagging RFI in $\chi^2$ data would be to set an absolute threshold for $\chi^2 / \mathrm{nDoF}$, and to flag data which exceed that threshold.
However, both the EoR0 and EoR1 fields have an overall Local Sidereal Time (LST) dependence in the $\chi^2$ values.
This makes it difficult to define an absolute $\chi^2 / \mathrm{nDoF}$ cutoff for flagging data, since what may be an elevated $\chi^2$ value for one observation can be the $\chi^2$ value of perfectly uncontaminated data for another.
Figure \ref{fig:LST_v_mean} illustrates this LST-dependence with the North-North polarizations of EoR1 observations.
Y-axis locations represent the mean of $\chi^2/\mathrm{nDoF}$ for each observation, and the error bars signify their standard deviations.
Points are color-coded by integer modified Julian date, such that data taken during the same 24-hour period are of the same color.
Different marker shapes represent different pointings of the MWA beam.

In the EoR1 data especially, there is a clear upward trend in each individual pointing's $\chi^2/\mathrm{nDoF}$ means.
This sub-pointing trend is likely due to bright celestial objects moving through the sidelobes of the individual tile beams, which are much less redundant between the tiles of the MWA than the beam's main lobe \citep{line_2018,chokshi_2021}; see \cite{choudhuri_2021} for a similar effect seen in simulations of HERA. 
Having the same bright sources cross our beam multiple times as the beam pointing is adjusted is natural in ``drift and shift'' observations such as these, where the primary beam pointing is adjusted to different angles at regular intervals throughout the observing period.

Taking the modified z-scores of the $\chi^2/\mathrm{nDoF}$ data greatly reduces their LST-dependence, as shown in Figure \ref{fig:LST_v_mean_z-score}, which represents the same data as Figure \ref{fig:LST_v_mean}.
Modified z-score for the $i$\textsuperscript{th} element of a dataset $x$ is calculated according to the following formula:
\begin{align}\label{z-score}
    z^{}_i = 0.6745\times\frac{(x^{}_i - \tilde{x})}{(\mathrm{MAD})}
\end{align}
where $\tilde{x}$ represents the dataset's median, and $\mathrm{MAD}$ is the median absolute deviation of the data from the median.
\begin{align}\label{MAD}
    \mathrm{MAD} = \widetilde{\left|x^{}_i - \tilde{x}\right|}
\end{align}
Here, the tilde ($^\sim$) operator has been used to signify ``take the median of the dataset.''
The factor of $0.6745$ is included to scale modified z-scores to have the same step size as standard z-score, i.e., if the standard z-score and modified z-score are both taken of the same perfectly Gaussian dataset, each integer step in modified z-score is the same size as a step in standard z-score, and corresponds to $1$ standard deviation of the dataset.

As opposed to the standard z-score, modified z-score uses medians instead of means.
As a result, the measure is less affected by outlier values, and more appropriate for datasets that have a mostly Gaussian distribution, aside from some high z-score outliers (in this case caused by RFI).

Modified z-score assumes that the data have a distribution that is close to Gaussian; this is a valid assumption for these data because, even though a $\chi^2$ distribution is not Gaussian, due to the central limit theorem, as nDoF increases, so does the distribution's resemblance to a Gaussian.
For the redundant sub-array of the MWA, as calculated with equation \eqref{dof}, $\mathrm{nDoF} = 2228$, which is more than enough to warrant making this simplification.

Because it almost completely eliminates LST-dependence of $\chi^2$, we base our RFI-flagging algorithm on the modified z-scores of $\chi^2/\mathrm{nDoF}$ generated by redundant calibration.

As shown in Figure \ref{fig:LST_v_mean_z-score}, most files have a mean z-score which is near 0.0 and a standard deviation of z-scores which is close to 1.0.
There are also files with noticeably higher mean z-scores, and wider standard deviations, which are almost always caused by strong RFI events.  However, rather than further analyze the distribution of $\chi^2$ values within a single observation, we turn to the time, frequency, and polarization information of the $\chi^2$'s, which help us localize and characterize the source of the interference.  


\subsection{Examples of \texorpdfstring{$\chi^2$}{chi2} and its Sensitivity to RFI}\label{examples}
\begin{figure*}[t!]
\caption{Waterfalls and histograms of the modified z-scores of $\chi^2/\mathrm{nDoF}$ of 6 two-minute observations. The waterfall plots, left, show the modified z-score of $\chi^2 / \mathrm{nDoF}$ as a function of time and frequency. Elevated values are indicative of RFI. To the right of each waterfall plot is the corresponding histogram, showing the distribution of the modified z-scores of the data. The first observation is RFI-free and shows a nearly Gaussian normal distribution, whereas the subsequent examples all contain RFI and have high-z-score outliers of varying degree. The second through fifth images all represent DTV events, while the sixth is a narrow-band event at 196.175 MHz.}
\label{fig:waterfalls_and_histograms}
\centering
\includegraphics[width=\textwidth]{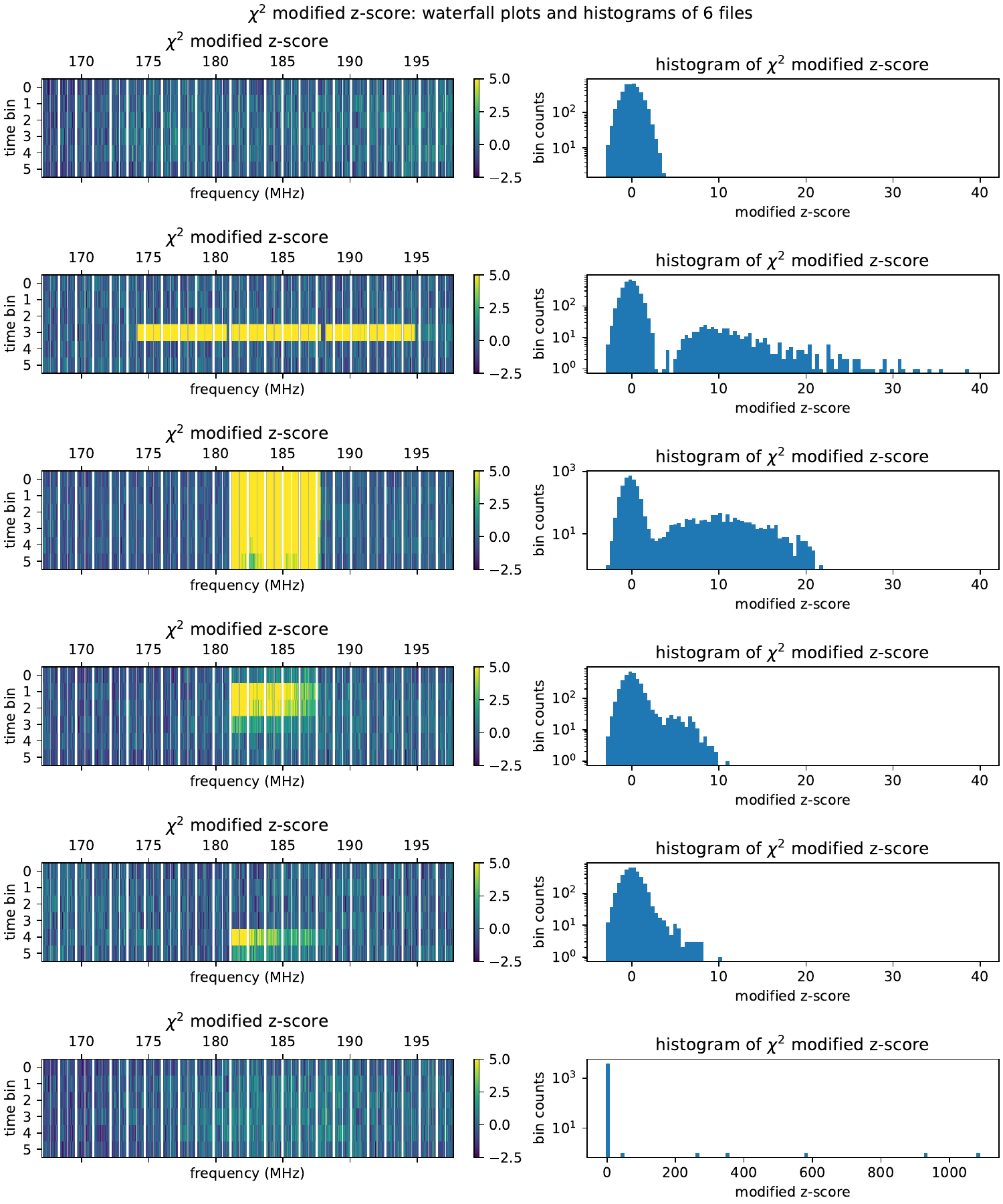}
\end{figure*}

To demonstrate the sensitivity of $\chi^2$ to RFI, we have included here a few $\chi^2$ time/frequency waterfall plots and histograms of the corresponding data in Figure \ref{fig:waterfalls_and_histograms}.

While the RFI-free file produces a distribution of $\chi^2$ that is close to Gaussian, the presence of RFI leads to outliers.
Included are four examples of DTV RFI of differing strengths.
Common to all four of them is a bimodal distribution of $\chi^2$, which generally remains below $z=40$, but, even in the example of the weakest RFI, have bins above $z=4$.

On the other hand, the RFI in the final waterfall plot of Figure \ref{fig:waterfalls_and_histograms} may be difficult to see, because it only spans one frequency-bin at 196.175 MHz.
It is worth noting, however, that the $\chi^2$ z-score values are exceptionally high for these six bins, as demonstrated in the histogram.

\subsection{RFI Flagging}\label{flagging}
RFI appearing in the $\chi^2$ data is flagged with a two-step process consisting of iterative modified z-score flagging followed by watershed flagging.

First, the modified z-scores of the $\chi^2$ values are calculated for the dataset.
Each point with a modified z-score greater than 4.0 is first masked out of the dataset. After high-z-score datapoints have been flagged, z-scores are recalculated without the outlier points, and the process is repeated until it produces no update in flags from its most recent iteration.

Starting from this preliminary set of flags, we follow the watershed RFI flagging algorithm described in Appendix A of \cite{HERA_flagging}.
Briefly, this flagging algorithm works on the assumption that RFI-contaminated bins tend to be contiguous, and flags every bin with a modified z-score greater than 2.0 which lies orthogonally adjacent to an already-flagged bin.
This process is iteratively repeated until the flags remain unchanged from the last iteration, thereby ``flooding'' the lower-z-score data which are directly connected to higher outliers.

This algorithm has to be slightly modified in order to handle the coarse-band flagging which is necessary for Phase II MWA data.
In general, when working with MWA Phase II data, within each 1.28 MHz, 32-channel coarse band, two channels on either edge of the band as well as the central channel are flagged and excluded from analysis due to data corruption arising from aliasing from the poly-phase filter bank (for the edges) and DC offsets (for the center) \citep{offringa_2015}.
These flags are apparent in the masked values in the waterfall plots of Figure \ref{fig:waterfalls_and_histograms}.  As a result, the function for finding ``adjacent'' bins has to be modified to ignore the coarse band edges.
When a bin being processed by the watershed algorithm is adjacent to a coarse band flag, the algorithm skips the flagged coarse feature and considers the next bin beyond it as adjacent to the flagged bins on the other side for ``flooding.''

\begin{figure}[ht!]
\caption{An example $\chi^2$ waterfall plot of an incompletely-flagged channel 7 DTV event. The top panel shows the calculated z-scores, while the bottom panel shows the derived flags in magenta.  This event was flagged using the algorithm described in Section \ref{flagging}, but the channel 7 DTV band is only partially flagged for any given time.}
\label{incomplete_flagging}
\centering
\includegraphics[width=\textwidth]{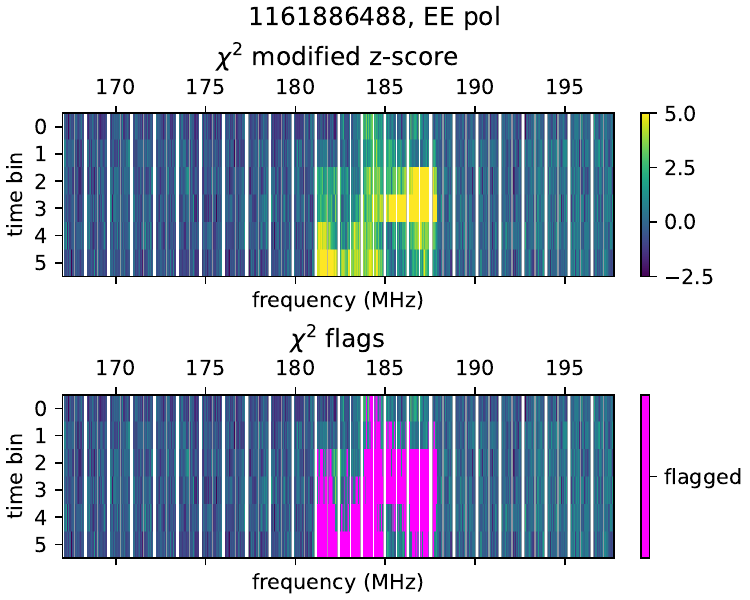}
\end{figure}

This two-step flagging algorithm is far from perfect.
Even after watershed flagging, there are datapoints which are clearly contiguously part of an RFI event which remain unflagged due to their low z-scores.
One such event is shown in Figure \ref{incomplete_flagging}.
While the two-step algorithm has nearly completely flagged contiguous points with high modified z-score, we can identify this RFI event as a DTV channel 7 broadcast due to its frequency band.
Using this deduction, we can assume \emph{a priori} that the RFI should occupy the entire channel 7 band, which is not fully flagged by this code.

In this work, we focus on demonstrating the potential for redundant calibration $\chi^2$ to detect RFI events.
Future work is likely to improve on the exact algorithm for calculating flags from the $\chi^2$ values themselves.
One might, for instance, explore averaging similar data together to reduce noise in $\chi^2$, since we can expect noise to reduce with averaging whereas consistent RFI will not.
Furthermore, the prospect of using a machine-learning approach to flag RFI events is especially interesting, based on the heuristic that they are easy for the human eye to identify, while still being difficult to algorithmically define.

\section{Other RFI-Flagging Algorithms: AOFlagger and SSINS}\label{other_algorithms}
In order to compare $\chi^2$ flagging with existing RFI flagging algorithms, we must obtain these algorithms' flags for the same dataset.

For the purposes of this work, we have compared the performance of $\chi^2$ flagging with AOFlagger as part of the MWA \verb|cotter| pipeline \citep{offringa_2015} and SSINS \citep{absolving_ssins}, run locally.

\subsection{AOFlagger}\label{aoflagger}
AOFlagger operates in three broad steps.
First, a high-pass filter is applied to the visibility amplitudes.
The filtered visibility amplitudes are then run through a straight-line detecting algorithm called SumThreshold, introduced in \cite{offringa_2010}, based on the observation that RFI often has sharp edges in the frequency-time domain.
Finally, the scale-invariant rank (SIR) operator, introduced in \cite{offringa_2012}, is applied to the SumThreshold-flagged visibilities, which acts to propagate those flags into adjacent contaminated bins.

AOFlagger flags are generated by \verb|cotter| (the processing pipeline in use by the MWA during Phase II operations), and included when downloading MWA data from the data download service ASVO.
The AOFlagger algorithm generates flags by baseline, so in order to visualize the flagging patterns for an entire individual observation, we average over the baseline axis and view the average occupation fraction for each frequency/time bin for a given polarization.

\subsection{SSINS}
SSINS is short for Sky-Subtracted Incoherent Noise Spectra, and is an RFI-flagging algorithm first introduced in \cite{absolving_ssins}\footnote{\url{https://github.com/mwilensky768/SSINS}}.
SSINS begins with ``sky-subtraction,'' whereby every visibility is subtracted from the next visibility adjacent in time, removing features that vary much more slowly than the two-second cadence of most MWA data, including the sky.
Once sky-subtraction has been applied to every baseline's visibilities, the data should consist of only noise and RFI.
The sensitivity to RFI is increased by forming an ``incoherent noise spectrum,'' or INS, of the sky-subtracted visibilities.
This is done by averaging the visibility amplitudes (discarding phase) over all baselines in the array.
One INS per polarization is produced.

SSINS uses a frequency-matched flagging algorithm to find features in the sky-subtracted incoherent noise spectrum.
This algorithm is aware of DTV-bands and will flag out the entire channel band if enough DTV RFI is detected in that channel.   
The flagging algorithm is also sensitive to narrow-band RFI and broad-band streaks.

SSINS version 1.4.7 was run locally on the downloaded data.
In order to reduce the number of spurious flags, the incoherent noise spectrum (INS) object was calculated with a 2\textsuperscript{nd} order polynomial fit for each frequency channel during mean subtraction.

\section{Results}\label{results}

\begin{figure}[ht]
\caption{The fraction of data flagged using $\chi^2$ across all observations as a function of frequency. The most commonly flagged RFI (in approximately 5 to 6\% of the data) is channel 7 DTV, as indicated by the large feature between 181 and 188 MHz. Also visible are slight amounts of flagging in channels 6 (174-181 MHz) and 9 (195-202 MHz). Narrow band RFI is apparent at 196.175 MHz and 194.455 MHz. Most other narrow-band peaks correspond to MWA coarse band edges, indicated by the shaded regions and are unlikely to be actual RFI. The slight increase in flagging at the lowest frequencies is due to spurious flagging of elevated $\chi^2$ stemming from the presence of the Galactic plane in the sidelobes of the primary beam (see Figure \ref{fig:galaxy}).}
\label{fig:flagging_fraction}
\centering
\includegraphics[width=\textwidth]{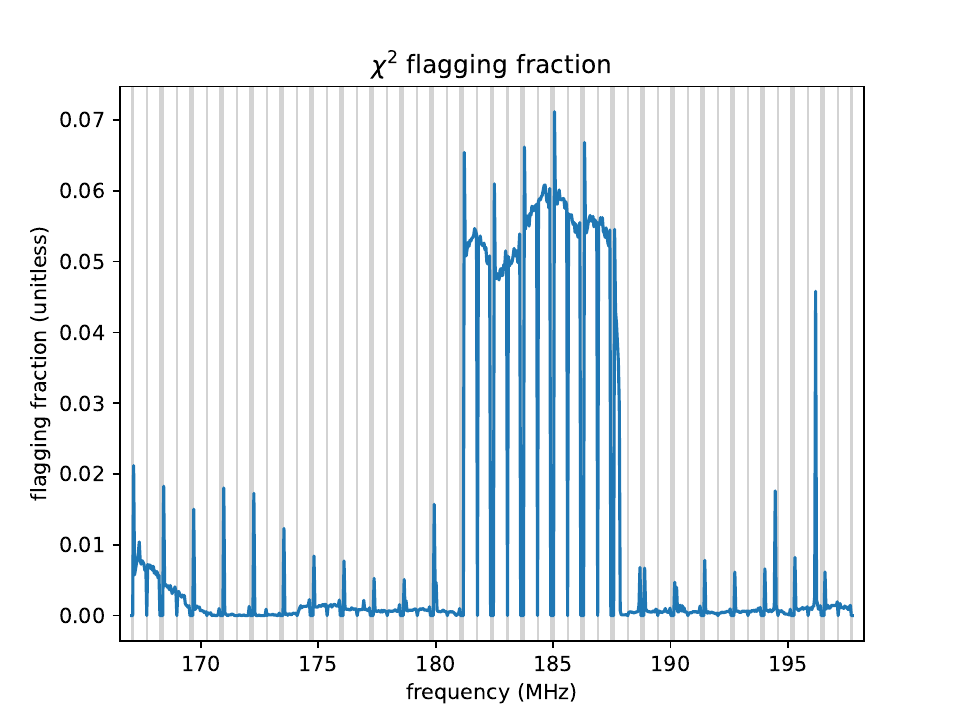}
\end{figure}

\subsection{Average flagging occupation fractions}\label{flagging_fraction}

When the $\chi^2$ flags are averaged across all of the files in the dataset, as visualized in Figure \ref{fig:flagging_fraction}, it becomes clear that RFI falling in the band allocated to Australia's digital television channel 7 (181-188 MHz) comprises the bulk of the RFI which is flagged.
Looking closely, we see that DTV channels 6 (174-181 MHz) and 9 (195-202 MHz)  are also visible as slight elevations in flagging fraction.

In addition to the strong peak at channel 7 frequencies, there are some noticeable narrow-band peaks.
Most of these narrow-band peaks in bins adjacent to the masked-off bins corresponding to the MWA correlator's coarse bands, which are widely known to contaminate data, as described in Section \ref{flagging}.
Coarse band flags are represented by the shaded zones in Figure \ref{fig:flagging_fraction}, and data from these frequency bins have been excluded from this plot.
Most of the observed narrow-band peaks in flagging fraction are adjacent to these flagged channels, which perhaps indicates that the data corruption extends beyond the commonly-flagged channels, although the flagging fraction is only $1-2\%$, so it follows that their $\chi^2$ values are not consistently elevated.
The cause of the elevated $\chi^2$ numbers that occasionally appear around the coarse band edges remains a topic for further investigation.

Of the narrow-band flagging peaks, two prominent examples are \emph{not} adjacent to coarse band edges.
These represent real narrow-band RFI, each contained within a single 40 kHz frequency-bin at central frequencies of 196.175 MHz and 194.455 MHz for the higher and lower peaks, respectively.

\begin{table*}[htbp] 
    \centering
    \begin{tabularx}{\textwidth}{ X | X | X | X | X | X | X | X | X } 
             & flagged by none & flagged by $\chi^2$ only & flagged by SSINS only & flagged by AOFlagger only & flagged by $\chi^2$ and SSINS only & flagged by $\chi^2$ and AOFlagger only & flagged by SSINS and AOFlagger only & flagged by all\\
        \hline
           number of observations & 1209 & 169 & 143 & 0 & 30 & 13 & 14 & 87 \\
        \hline
            \% of total  & 72.61\% & 10.15\% & 8.59\% & 0.00\% & 1.80\% & 0.78\% & 0.84\% & 5.23\%\\
        \hline
            \% of flagged  & n/a & 37.06\% & 31.36\% & 0.00\% & 6.58\% & 2.85\% & 3.07\% & 19.08\%\\
    \end{tabularx}
    \caption{Number of files flagged in the DTV channel 7 band for $\chi^2$, SSINS, and AOFlagger.}
    \label{tab:table}
\end{table*}

\begin{figure}[ht]
\caption{An example of an observation with elevated low-frequency NN $\chi^2$ values due to the presence of the Galactic plane in the sidelobes of the MWA primary beam. In the corresponding histogram, we can see that the distribution deviates from a Gaussian normal. Observations like this (all corresponding to the -3 pointing with the instrument phased to EoR0) cause the uptick in flagging at very low frequencies in Figure \ref{fig:flagging_fraction}.}
\label{fig:galaxy}
\centering
\includegraphics[width=\textwidth]{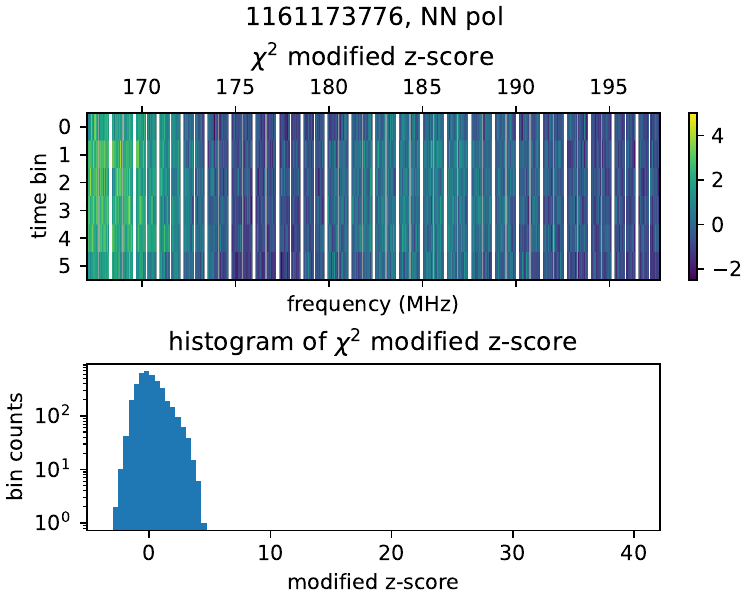}
\end{figure}

\begin{figure}[ht]
\caption{A Venn diagram comparing the percentage of DTV events flagged by the $\chi^2$, SSINS, and AOFlagger algorithms. Most of the events flagged by AOFlagger are detected by either $\chi^2$ or SSINS, but there are a significant number which are flagged by $\chi^2$ and missed by SSINS, and vice-versa. The circles in this diagram are to scale, although overlaps are not.}
\label{fig:venn}
\centering
\includegraphics[width=1.0\textwidth]{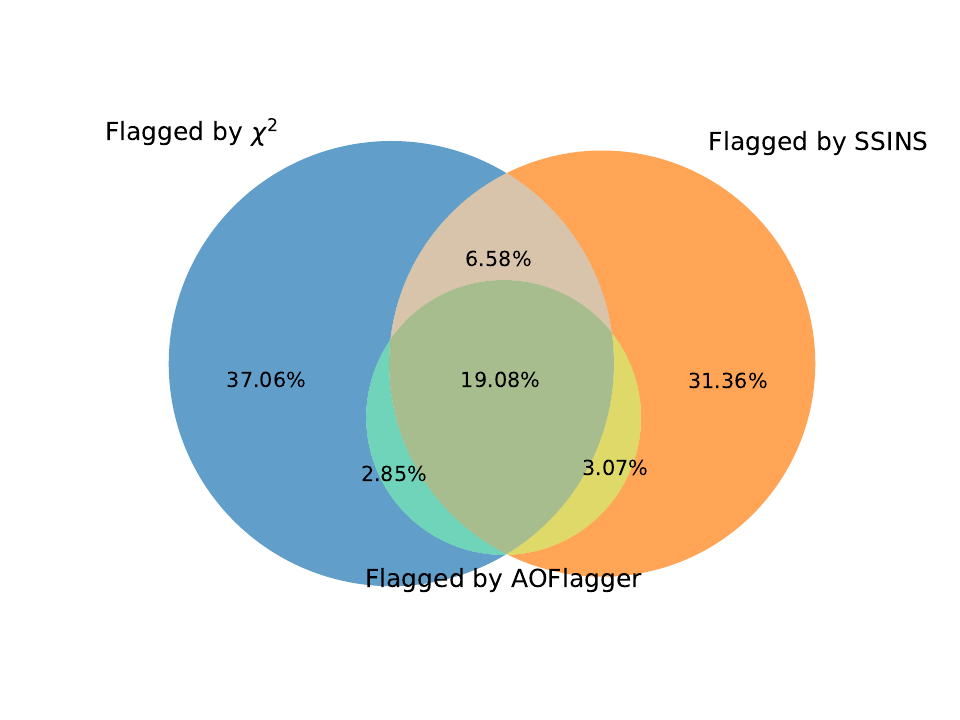}
\end{figure}

The last obvious feature of Figure \ref{fig:flagging_fraction} is a small increase in flagging at low frequencies, which is due to the spurious flagging of low-frequency bins whose $\chi^2$ values are elevated --- not by RFI, but by the presence of the Galactic plane in the far sidelobes of the primary beam.
Affected observations, like the one visualized in Figure \ref{fig:galaxy}, are uniformly from gridpoint number $\minus 3$ of the grid pointing scheme used in \cite{beardsley_2016}, with az $= 90^{\circ}$, el $= 69.1655^{\circ}$ --- one of the more extreme pointings of the instrument used to observe the EoR0 field which appear in this dataset.
The affected data were taken when the Galactic anti-center was still above the horizon, where it could be picked up by sensitivity in the sidelobes of the beam.
Although using modified z-score for flagging decreases the number of spurious flags in this pointing, it does not reduce them to zero.
This results in the uptick in $\chi^2$ flagging at the lowest frequencies observable by the MWA.  Future work looking to apply RFI flags generated from $\chi^2$ values to MWA data may need to handle the analysis of this pointing separately.

\begin{figure*}[ht!]
\caption{A comparison of $\chi^2$, SSINS, and AOFlagger performance on flagging DTV channel 7, the most commonly-detected source of RFI in our dataset. The x-axis represents the local time of the observation in UTC, while each vertical panel represents observations from a different date.  Within a panel, each observation is represented by three circles: the top row (blue circles) is our $\chi^2$ method, the middle row (orange circles) is SSINS, and the bottom row (green circles) is AOFlagger.  Filled circles indicate that RFI was detected by an algorithm in the channel 7 band for that file, whereas empty circles indicate that the algorithm did not detect any RFI in that observation file. ``Detected'' meaning, in the case of $\chi^2$ and AOFlagger, that $>5\%$ of possibly contaminated bins were flagged, and in the case of SSINS, that the match filter identified DTV channel 7 in this file. $\chi^2$ seems well-suited to detecting long-duration DTV events which are picked up only sporadically by SSINS and AOFlagger. Meanwhile, SSINS detects isolated events which are not detected by $\chi^2$ and, often, AOFlagger.}
\label{fig:dots}
\centering
\includegraphics[width=\textwidth]{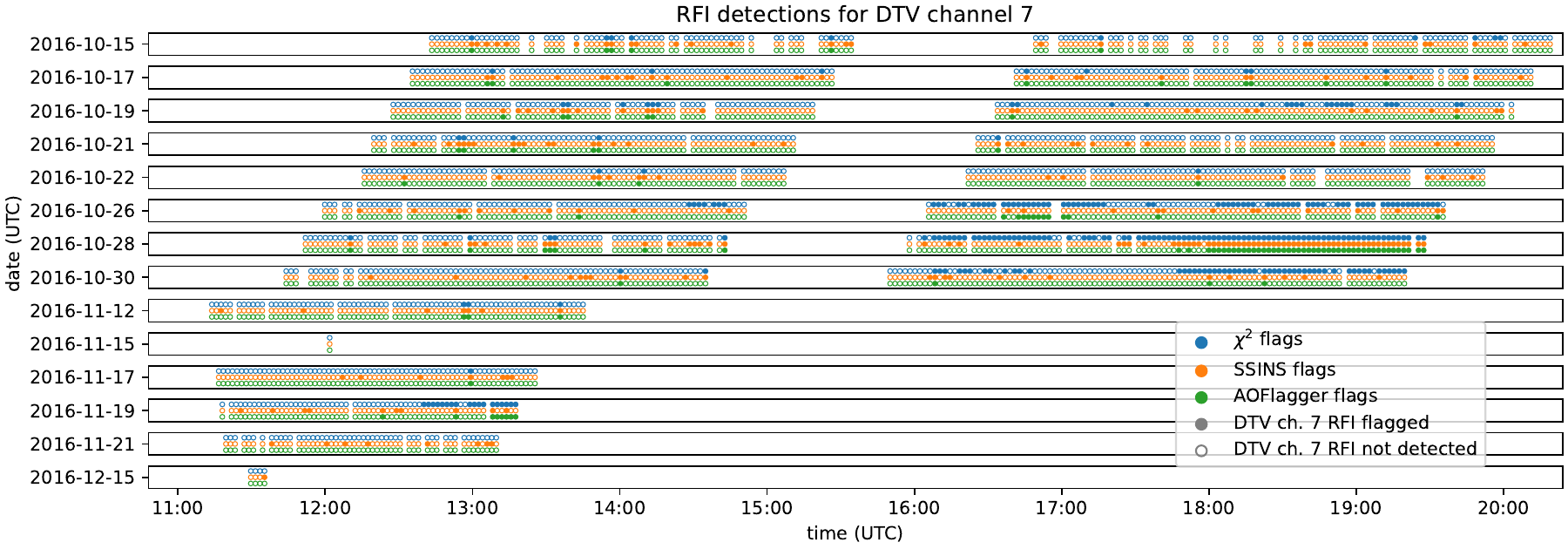}
\end{figure*}

\subsection{Comparing \texorpdfstring{$\chi^2$}{chi2} flagging with SSINS and AOFlagger}
The most common type of RFI detected in this dataset was channel 7 DTV, which is broadcast between 181 MHz and 188 MHz.
This section focuses on flagging RFI when only this band is considered.
Other types of RFI are discussed in Section \ref{discussion_and_conclusion}.

In Figure \ref{fig:dots}, a filled dot indicates that channel 7 DTV RFI was detected in a particular observation, whereas an empty dot indicates that channel 7 DTV RFI was not detected in that observation.
In order to qualify as ``detected'' by $\chi^2$ or AOFlagger, a file must clear a threshold value of 5\% of channel 7 bins being flagged by the algorithm in question, as well as verifying that the DTV channel 7 band is more densely flagged than the file as a whole (which distinguishes channel-7-only RFI from broad-band events).
To qualify as ``detected'' by SSINS, a file must be tagged as containing channel 7 RFI by SSINS's match filter.
Although Figure \ref{fig:flagging_fraction} indicates that between 5\% and 6\% of the data are flagged for channel 7 RFI by $\chi^2$, as indicated in Table \ref{tab:table}, roughly 27.4\% of the two-minute observation \emph{files} are flagged for this type of RFI.
That is to say, 27.4\% of files have 5\% or more of the DTV channel 7 band flagged (if the detection was with $\chi^2$ or AOFlagger) or was tagged by SSINS's match filter as containing DTV channel 7, but if all of the data is considered in bulk, between 5\% and 6\% of the total data are contaminated by channel 7 DTV (according to $\chi^2$-based flagging).

When the flagging fraction of the channel 7 band is compared across algorithms and plotted versus time, as in Figure \ref{fig:dots}, we observe that there are certain DTV events which are flagged by $\chi^2$ but not the other algorithms, and that the same could be said for SSINS.
The number of files flagged in the channel 7 band for each algorithm is summarized in Table \ref{tab:table} and in the Venn diagram in Figure \ref{fig:venn}.

While every event caught by AOFlagger was also found by either SSINS, $\chi^2$, or both, there is less overlap between the events flagged by SSINS and $\chi^2$.
Of all events flagged by $\chi^2$, about 56.5\% are unflagged by other algorithms, and of all the events flagged by SSINS, roughly 52.2\% are unflagged by other algorithms.
This indicates that the SSINS and $\chi^2$ algorithms are both suited to spotting exclusive categories of events, whereas there is a significant overlap between AOFlagger and either SSINS or $\chi^2$.

Using only these three algorithms, it is impossible to determine whether there are events which are missed by \emph{all three}; it is of course possible that a novel algorithm will be able to detect RFI events invisible to $\chi^2$, SSINS, and AOFlagger.
However, for the data and methods we have, it seems that using a combination of $\chi^2$ and SSINS will flag all detectable RFI, and the addition of AOFlagger does not measurably add to our sensitivity.

\section{Discussion and Conclusion}\label{discussion_and_conclusion}
\subsection{RFI classification}
Several distinct types of RFI can be seen in $\chi^2$ data.
The most common type of RFI is digital television (DTV) signals, usually in Australian channel 7 (181-188 MHz), although channel 9 (195-202 MHz) is sometimes also present.
Channels 6 (174-181 MHz) and 8 (188-195 MHz) are the rarest to see, usually only visible in very strong DTV RFI events which contain 7 and 9 as well.
Channel 7 is flagged in roughly 5-6\% of the data, channels 6, 8, and 9 are all flagged in less than 1\% of the data.

We also observe broad-band, short-duration RFI which is visible through most of the frequencies observed by the MWA.
In general, SSINS seems much more sensitive to these events, which seem difficult to flag using $\chi^2$.

Finally, we can see long-duration, narrow-band RFI which fits entirely into a single 40 kHz frequency-bin at central frequencies of 196.175 MHz, and 194.455 MHz, as discussed in section \ref{flagging_fraction}.
An example event is visualized in Figure \ref{fig:1161526216} in \ref{appendix}.

\begin{figure*}[ht]
\caption{The same as Figure \ref{fig:dots}, except for the 196.175 MHz narrow-band RFI signal (as opposed to channel 7 DTV). 
 This signal appears to be periodic with a period of around 105 minutes.}
\label{fig:narrow_band}
\centering
\includegraphics[width=\textwidth]{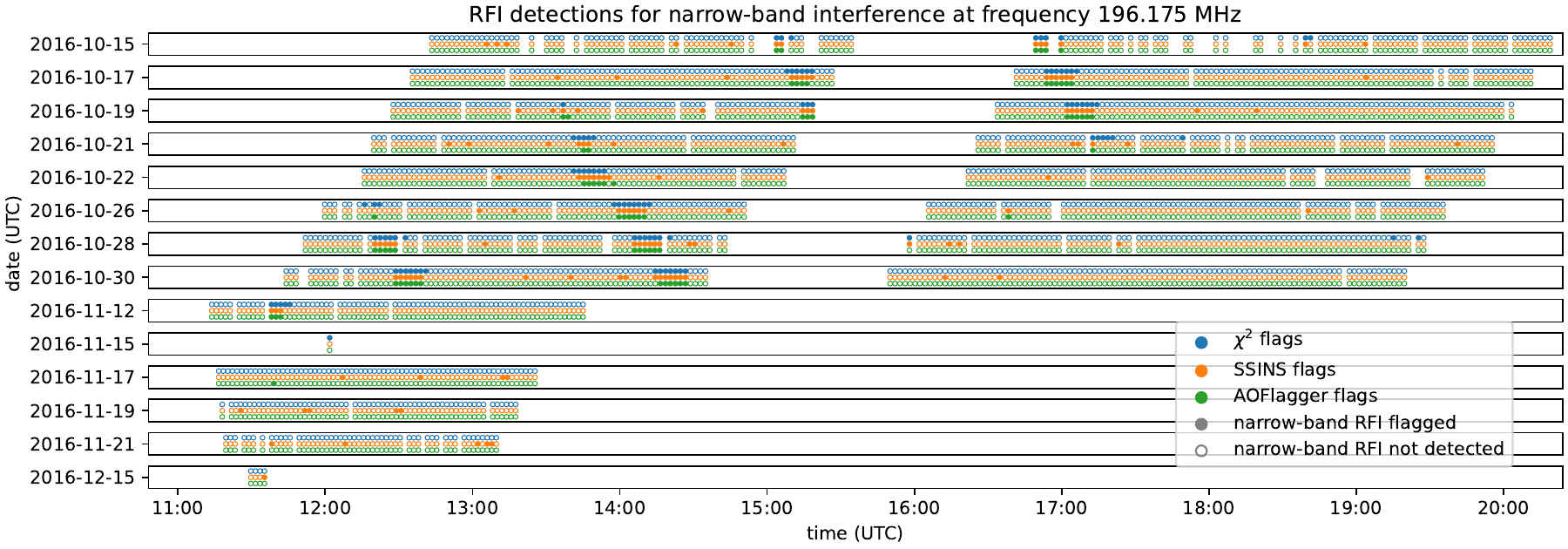}
\end{figure*}

When flagging occupancy for the 196.175 MHz narrow-band RFI is plotted against time, as in Figure \ref{fig:narrow_band}, it seems to represent a repeating signal with a period of around 105 minutes.
The source of this radiation is unclear, although the 105 minute period does fall in a range of periods typical for low Earth orbit satellites.
Further investigation is needed to track down this signal's provenance.

\subsection{Comparing events flagged by different algorithms}
Referring to Figure \ref{fig:dots}, we observe a few patterns:
\begin{itemize}
\item Observations flagged by only $\chi^2$ tend to occur clustered in time, and are more common in the data taken later at night (which correspond to EoR1 observations).
    \begin{itemize}
    \item an example of an observation that was flagged by $\chi^2$ alone, and was a part of one of these late-night clusters is Figure \ref{fig:chisq_only} (this and other examples in this section may be found in \ref{appendix}). This observation is from the second half of the night on 2016-10-26, around 18:15 UTC.
    \end{itemize}
\item SSINS flags also cluster in time somewhat, but less consistently than $\chi^2$ flags, and the clusters seem shorter in duration on average (consistent with the time clustering properties found in \citep{wilensky_et_al_2023}).
    \begin{itemize}
    \item Figures \ref{fig:1161870384} and \ref{fig:1161870256} are part of a short cluster of SSINS flags, at around 14:30 on 2016-10-30.
    \item While Figure \ref{fig:1161870256} is a bright signal in SSINS, it is brief enough to be reduced in strength by time-averaging. Figure \ref{fig:1161870384}, on the other hand, is much weaker when visualized in SSINS, even though it has a longer duration. It is unclear why the event in Figure \ref{fig:1161870384} was visible to SSINS but not to $\chi^2$.
    \end{itemize}
\item There are also many observations flagged by SSINS alone which are isolated from other flags in time.
    \begin{itemize}
    \item An example of an event which is isolated from other detections may be found at Figure \ref{fig:1162986688}. Like Figure \ref{fig:1161870256}, it is bright and brief. It occurred on 2016-11-12 shortly after 11:45.
    \end{itemize}
\item AOFlagger flags for this dataset overlap entirely with SSINS flags, $\chi^2$ flags, or both.
    \begin{itemize}
    \item It seems as though AOFlagger is not uniquely sensitive to any of the RFI in this dataset.
    \item An event which was flagged by AOFlagger and $\chi^2$ but not seen by SSINS can be seen at Figure \ref{fig:1163596672}. This event occurred just after 13:15 on 2016-11-19.
    \end{itemize}

\end{itemize}

In general, it seems that $\chi^2$ is more consistent in finding longer-timescale RFI than SSINS, and conversely, SSINS seems better at detecting short-timescale RFI.
An example of the former can be seen in figure \ref{fig:chisq_only}, and an example of the latter in Figure \ref{fig:1161870256}.
Given that the first step to processing the $\chi^2$ values of MWA data is to time-average the files, and this would have the tendency to ``wash out'' shorter-timescale events, this pattern makes sense.  Conversely, since SSINS relies on time-differencing to remove slowly varying signals, it is perhaps not surprising that it misses events with long durations. 

SSINS also sees certain wide-band, short duration RFI, termed ``streaks'', which are not as visible to $\chi^2$.
An example is Figure \ref{fig:1160575136}.

\subsection{\texorpdfstring{$\chi^2$}{chi2} and RFI}
In Section \ref{RFI_identification}, we discussed how $\chi^2$ is calculated and how it could be influenced by RFI.
We came to the conclusion that the heightened $\chi^2$ seen with RFI are not due to an underestimation of the noise variance $\sigma^2$, but instead due to genuinely less redundant visibilities between nominally redundant baselines.

RFI detected by $\chi^2$ has in the past been hypothesized to be locally produced, which would lead to certain antennas detecting the signal more strongly than others, thus increasing the $\chi^2$ metric.
In fact, in our analysis, it became clear that the RFI picked up by $\chi^2$ does not seem to be associated with local transmitters, as it doesn't peak in particular areas of the array for individual observations --- when $\chi^2$ is elevated for a certain observation, it is increased for every antenna in the array.

We offer the following theory for the elevated $\chi^2$ values we observe.  One source of non-redundancy in the MWA is the antenna-to-antenna variation of the the primary beam.  Observationally, the MWA beams are generally consistent in the main lobe, but the sidelobes exhibit more variability \citep{line_2018,chokshi_2021}.  RFI from sources in the sidelobes (e.g. signals propagated over the horizon via tropospheric ducting or reflections from airplanes far away from zenith) would appear with different strengths in the signals from different antennas, leading to more observed disagareement in nominally redundant visibilities and enhanced $\chi^2$ values.

As a corollary to this theory, we might expect this $\chi^2$ flagger to have a lower success-rate when RFI is located near the zenith, i.e., within the main lobe, since we expect significantly more agreement between redundant baselines.  The RFI events detected by SSINS but not $\chi^2$ could potentially be reflections off airplanes or other objects transiting near zenith, a detection-case which SSINS has been shown to handle well \citep{absolving_ssins}.  In order to verify this hypothesis and its implications, direct imaging of RFI caught by a $\chi^2$-based flagger is likely necessary.
This is an avenue for future work on this topic.


\subsection{Conclusion}\label{conclusion}
RFI detection and flagging remain a formidable challenge in the realm of 21 cm Epoch of Reionization cosmology.
Without an algorithm (or suite of algorithms) which can detect and excise RFI at a sensitivity heretofore unachieved, instruments like the MWA and HERA may be unable to deliver an accurate power spectrum of the Epoch of Reionization's 21 cm radiation.

In this work, we have presented $\chi^2$ from redundant calibration as a new approach to flagging RFI in data from the MWA.
We have demonstrated the effectiveness of this method and compared its efficacy at flagging different kinds of RFI than AOFlagger and SSINS.

$\chi^2$ flagging as implemented here seems effective at flagging the DTV RFI events which are all too common at the MWA site.
In the realm of long-duration, weak DTV events, $\chi^2$ finds RFI that's missed by the current state-of-the-art algorithm SSINS.
However, $\chi^2$ should not be taken as a replacement for or improvement on SSINS; in fact, the algorithms seem quite complementary, each being able to detect RFI that remains invisible to the other.



There are many potential avenues for future work using this tool.
The flagging algorithm based on the modified z-scores of $\chi^2/\mathrm{nDoF}$ could be improved with features like those found in the SSINS flagging suite, such as applying a threshold to fully flag a DTV channel when sufficient RFI has been detected within it.
Furthermore, a machine-learning approach to flag generation could potentially be sensitive to even weaker and more tenuous events, which are visible to the human eye but difficult to pick out algorithmically.

In the course of this work, a repeating narrow-band RFI source was identified at 196.175 MHz with a period of around 105 minutes.
Identifying the source of this RFI is another opportunity for further work.

Producing a cosmological 21\,cm power spectrum using data screened with $\chi^2$ RFI flagging alongside SSINS is another near-term goal.
Furthermore, it would be instructive to investigate whether this hypothetical power spectrum still carries the statistical footprint of ultra-faint RFI as described in \cite{wilensky_23}.

That the $\chi^2$ values produced by redundant calibration are sensitive to weak RFI is a happy accident, which will hopefully become useful in improving the quality of the data that goes into serious analyses in the future.
This work represents the first step in bringing this promising new method into the pipelines of new 21 cm data-processing efforts.

\section*{Acknowledgments}
Thank you to Bryna Hazelton, Miguel Morales, Michael Wilensky, and Pyxie Star for their support, encouragement, and handy SSINS tips.
Thank you also to our Cal-Bridge summer students Alexander Hawksley and Hal France for their initial investigations with $\chi^2$ RFI detection.

The authors would also like to acknowledge the developers of pyuvdata and \verb|hera_cal|.
These software projects were indispensable to this work.

This scientific work uses data obtained from Inyarrimanha Ilgari Bundara / the Murchison Radio-astronomy Observatory. We acknowledge the Wajarri Yamaji People as the Traditional Owners and native title holders of the Observatory site. Establishment of CSIRO's Murchison Radio-astronomy Observatory is an initiative of the Australian Government, with support from the Government of Western Australia and the Science and Industry Endowment Fund. Support for the operation of the MWA is provided by the Australian Government (NCRIS), under a contract to Curtin University administered by Astronomy Australia Limited. This work was supported by resources provided by the Pawsey Supercomputing Research Centre with funding from the Australian Government and the Government of Western Australia.

This work was conducted at Brown University. Brown University is located in Providence, Rhode Island, on lands that are within the ancestral homelands of the Narragansett Indian Tribe. The Narragansett Indian Tribe, whose ancestors stewarded these lands with great care, continues as a sovereign nation today. We commit to working together to honor our past and build our future with truth.

\paragraph{Funding Statement}

This research was supported by grants from the US National Science Foundation award IDs 	1907777 and 2106510.  

\paragraph{Competing Interests}

None.

\paragraph{Data Availability Statement}
Data are available to download at ASVO, \url{https://asvo.mwatelescope.org/}.
For OBSIDs corresponding to this dataset, and access to the $\chi^2$ flagging code, please contact the first author directly.


\printendnotes

\bibliography{biblio}


\appendix
\section{The effects of time-averaging on \texorpdfstring{$\chi^2 / \mathrm{nDoF}$}{chi2 per nDoF}}\label{effects_of_time_averaging}
In an initial analysis, the data were processed with \verb|hera_cal| without the additional nine-fold time averaging step described in Section \ref{preprocessing}. The result was that the $\chi^2$ values were noisy and elevated in regions where there is no RFI.

\begin{figure}[ht!]
\caption{A single time-slice of $\chi^2 / \mathrm{nDoF}$, generated by hera\_cal without downsampling in time first (top) and after downsampling in time (bottom). The result without downsampling is noisy and difficult to process to identify RFI. After downsampling, noise in the $\chi^2$ metric is significantly reduced, and the RFI is visibly elevated over the noise floor.}
\label{fig:noisy_chisq}
\centering
\includegraphics[width=\textwidth]{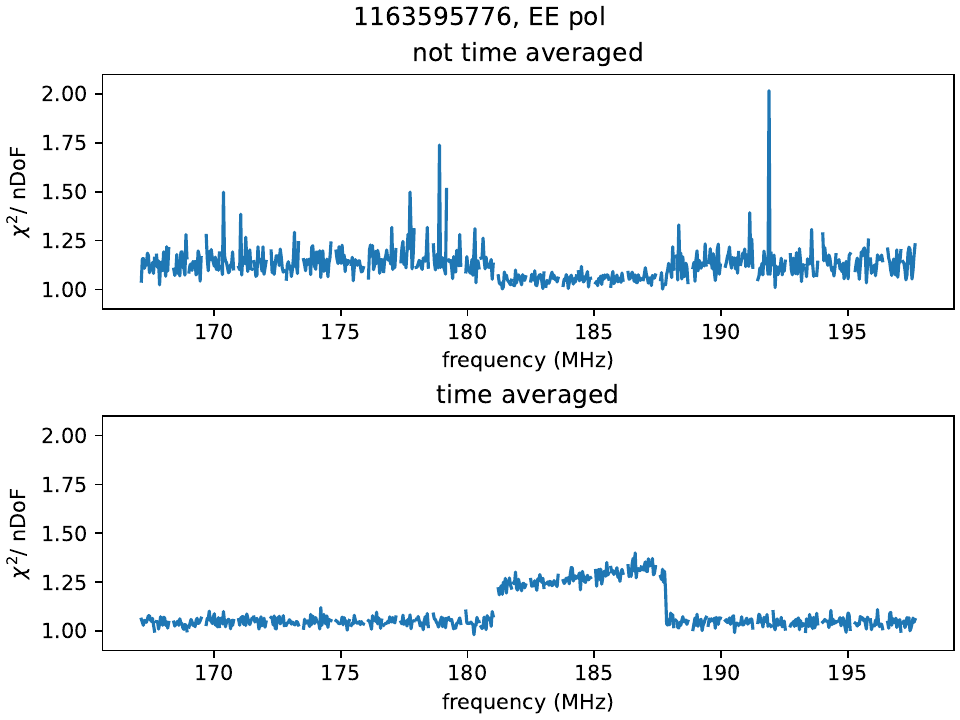}
\end{figure}

The upper plot in Figure \ref{fig:noisy_chisq} is a single time-slice of the $\chi^2 / \mathrm{nDoF}$ generated by \verb|hera_cal| from a file that was not first downsampled in time. 
This file is known to have DTV channel 7 RFI present, visible between 181 MHz and 188 MHz.
Notably, the presence of RFI increases the signal-to-noise level in those frequencies, and so there is much less noise propagated through to $\chi^2$ in that band.

The lower plot of Figure \ref{fig:noisy_chisq}, on the other hand, is also a single time-slice of the $\chi^2 / \mathrm{nDoF}$ generated by \verb|hera_cal| from the same file at approximately the same time, when the data had been time-averaged before calibrating.
Files are time-averaged from a cadence of one visibility every 2 seconds to one visibility every 18 seconds, which we would expect to lower the noise amplitude by a factor of $\sqrt{9} = 3$.
This plot showcases the lower noise levels afforded by time-averaging, and the stark contrast between frequencies with and without RFI.

\subsection{Weighting \texorpdfstring{$\chi^2$}{chi2} by NSamples to correct noise discrepancy}
When pyuvdata downsamples in time, it does not include flagged data in the calculated average.
Furthermore, when working with MWA data, it is good practice to flag a few times at the beginning and end of the file, which often contain bad data.
Noise is reduced by time-averaging, so we observe that $\chi^2$ generated from MWA data which have been flagged at the beginning and end of the file in this manner have an elevated noise-floor.

We can correct for this discrepancy by multiplying each $\chi^2$ value by the corresponding element in pyuvdata's NSample\_array, which tracks the fraction of unflagged data which were averaged into a given time/frequency bin.

When averaged, noise is decreased by a factor of $\sqrt{N}$, where $N$ is the total number of data-points averaged together.
Because $\sigma^2$ scales with the square of the noise level, and forms the denominator of equation \eqref{chisq}, we expect $\chi^2$ to scale inversely with $N$.
It is sufficient, then, to normalize the noise level by multiplying $\chi^2$ by the corresponding element in the UVData object's NSample\_array.

\section{redcal\_run() parameters for use on MWA data}\label{redcal_run}
A few custom parameters are necessary to make \verb|hera_cal| work with MWA data.
The parameters we used were \begin{verbatim}max_dims=3, oc_conv_crit=1e-10, oc_maxiter=2000,
    check_every=10, check_after=500,
    ant_z_thresh=100.0, gain=0.3\end{verbatim}

A few of these warrant further explanation -- \verb|max_dims=3| is absolutely necessary, because the spatial separation between the two hexagons introduces an extra dimension to the degenerate subspace inherent to redundant calibration solutions, corresponding to an arbitrary phase separation between the hexagons, in addition to the three other degenerate parameters, which correspond to the tip/tilt of the array and an overall phase. 
Setting \verb|ant_z_thresh=100| eliminates the software's throwing out of high-z-score antennas when calculating solutions, because we are \emph{not} planning to use the gain solutions for calibration. 
The remaining parameters are the result of some by-hand fine-tuning, but could probably be improved with a more systematic parameter-space search.

\section{A selection of interesting RFI events}\label{appendix}

In the course of this research, we encountered many interesting individual RFI events.
In this appendix, we have included a selection of particularly illustrative or interesting events.

\begin{figure*}[ht]
\caption{A typical example of an observation with RFI in DTV channel 7 which is detectable with $\chi^2$, and not with other flagging algorithms. This observation is a part of a long string of DTV detections seen by $\chi^2$ alone.}
\label{fig:chisq_only}
\centering
\includegraphics[width=\textwidth]{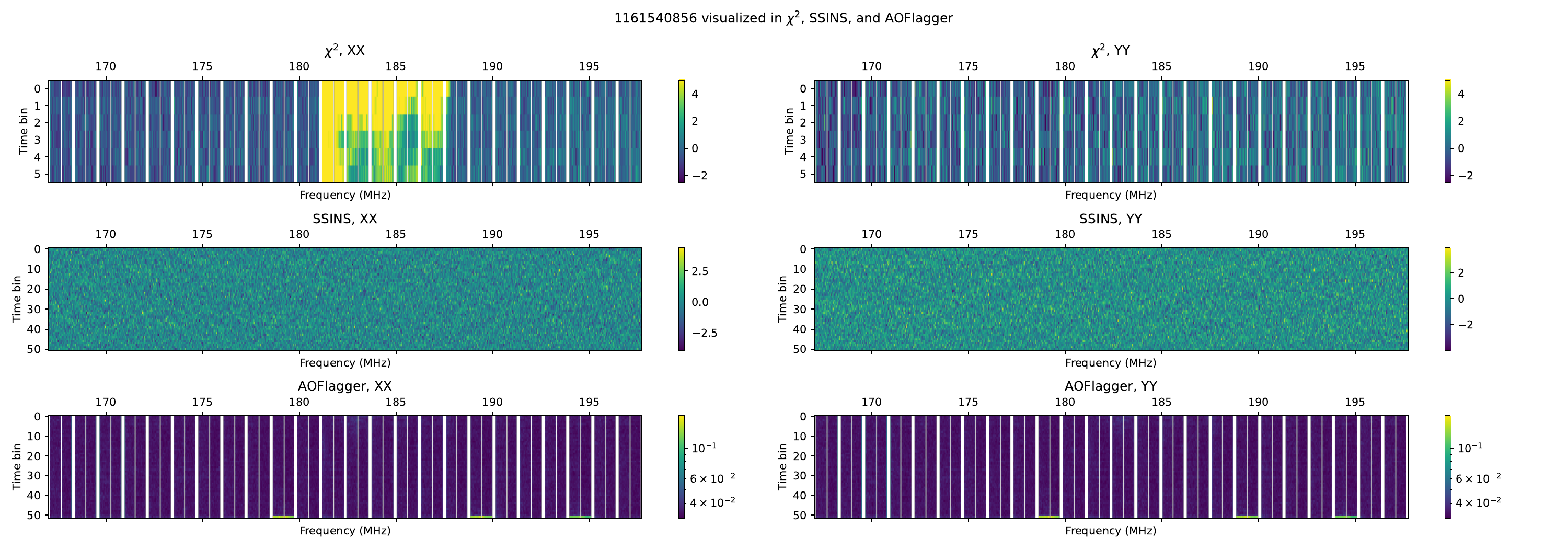}
\end{figure*}

\begin{figure*}[ht]
\caption{This SSINS DTV detection occurs as part of a small cluster of files flagged by SSINS alone. The event is also visible in AOFlagger, and a hint of it is visible in $\chi^2$, but it did not reach the threshold for being listed as being ``detected'' by any algorithm besides SSINS.}
\label{fig:1161870384}
\centering
\includegraphics[width=\textwidth]{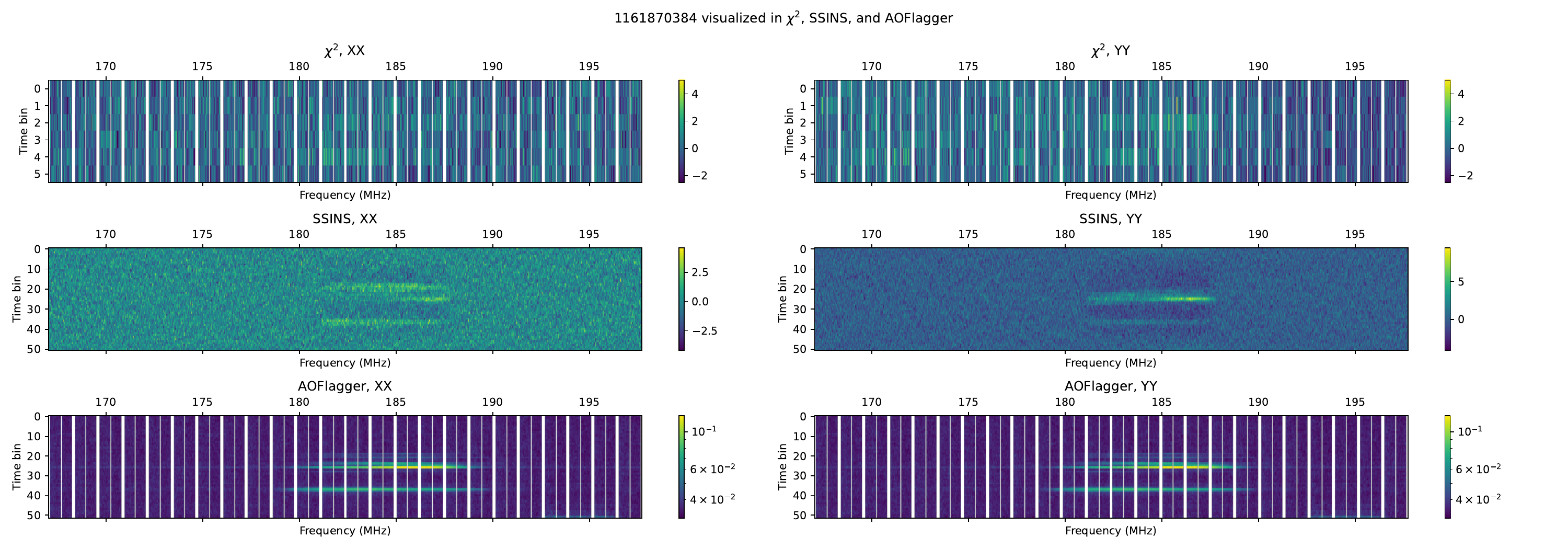}
\end{figure*}

\begin{figure*}[ht]
\caption{This SSINS DTV detection occurs as part of the same small cluster of SSINS detections as Figure \ref{fig:1161870384}. The event is also visible in AOFlagger, but it did not reach the threshold for being listed as being ``detected'' by AOFlagger or $\chi^2$.}
\label{fig:1161870256}
\centering
\includegraphics[width=\textwidth]{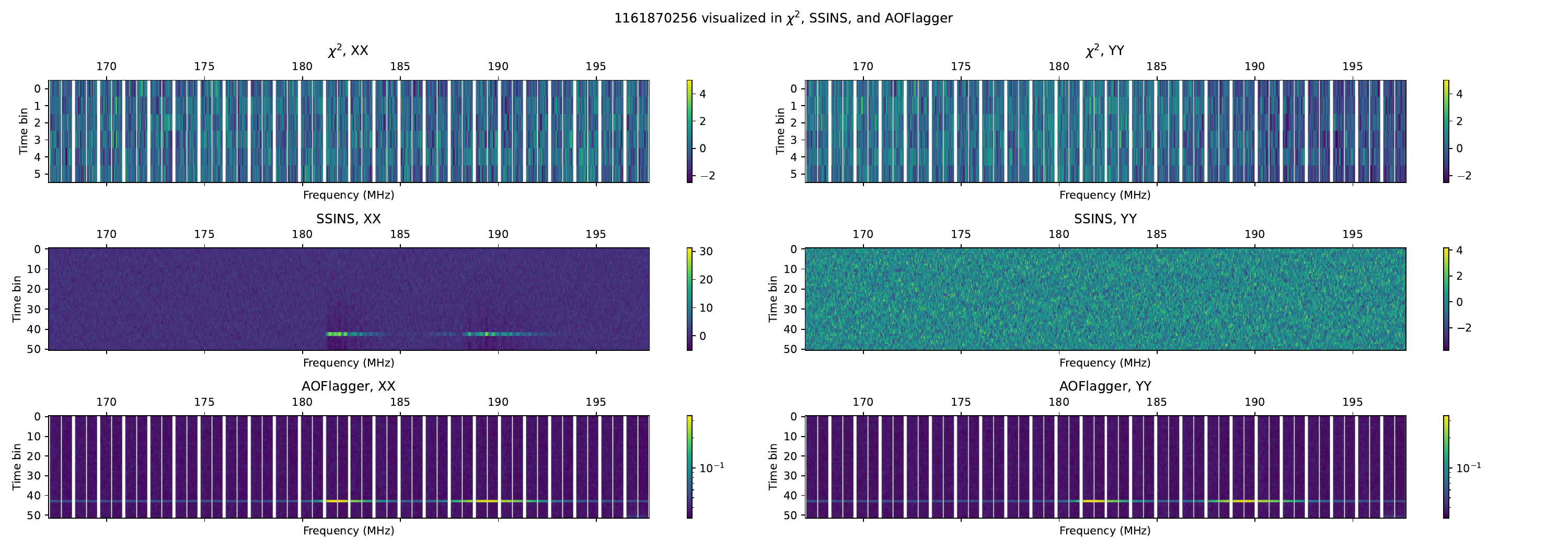}
\end{figure*}

\begin{figure*}[ht]
\caption{This SSINS DTV detection occurs isolated in time from other RFI events.}
\label{fig:1162986688}
\centering
\includegraphics[width=\textwidth]{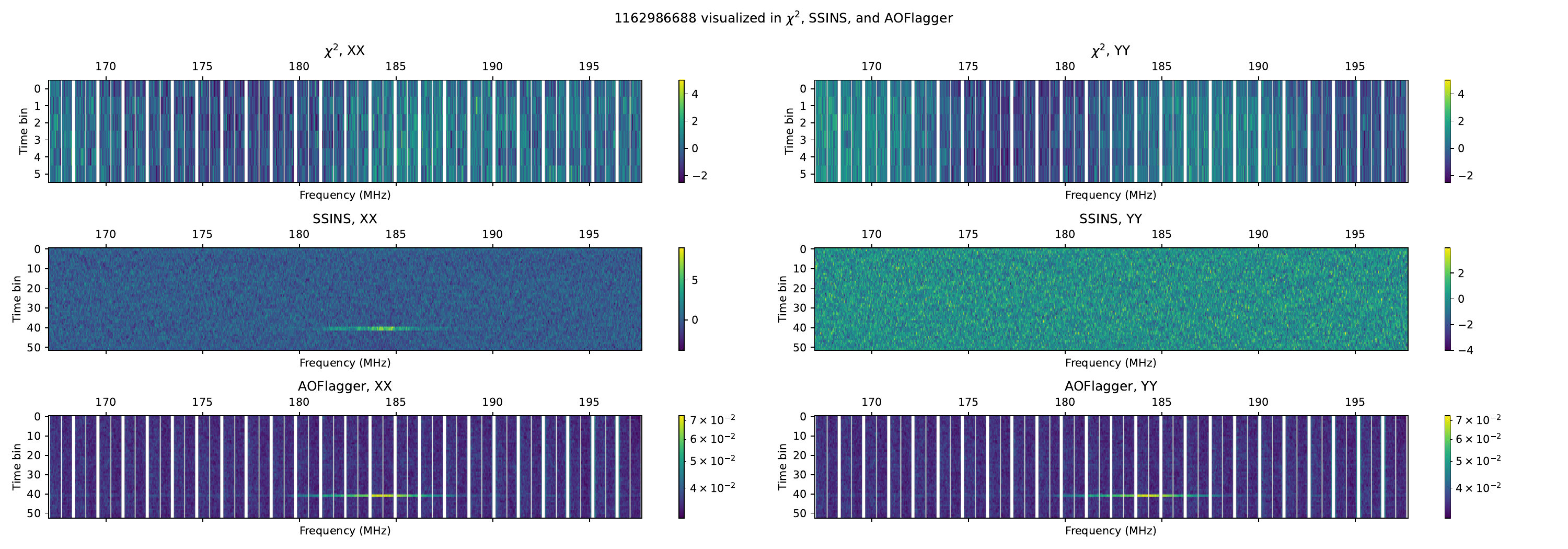}
\end{figure*}

\begin{figure*}
    \centering
    \includegraphics[width=\textwidth]{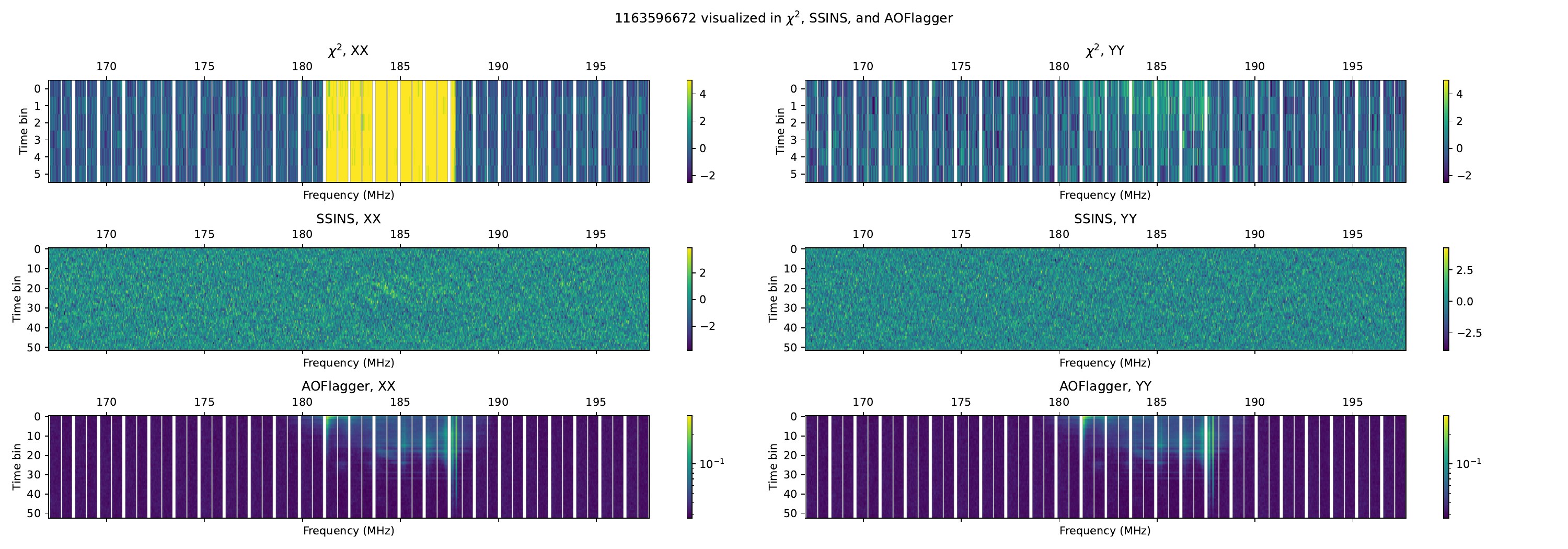}
    \caption{This DTV channel 7 event was detected by $\chi^2$ and AOFlagger, but missed by SSINS.}
    \label{fig:1163596672}
\end{figure*}

\begin{figure*}[ht]
    \centering
    \includegraphics[width=\textwidth]{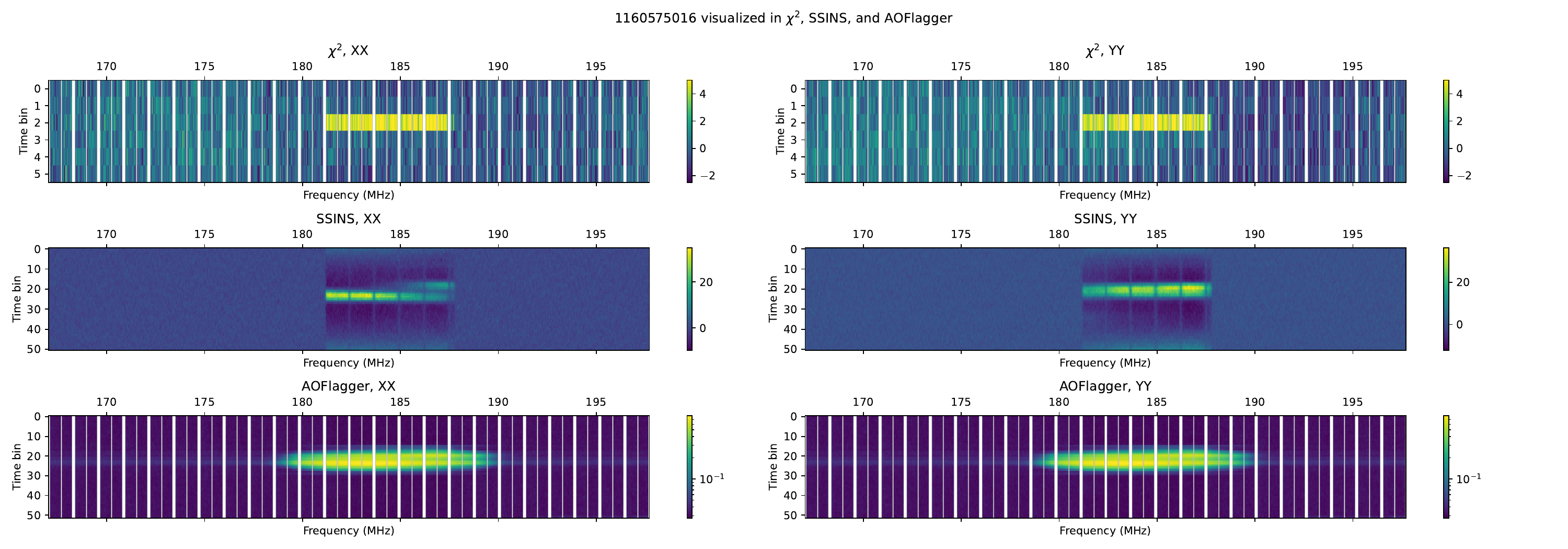}
    \caption{An example of DTV channel 7, detected by all three algorithms.}
    \label{fig:1160575016}
\end{figure*}

\begin{figure*}[ht]
    \centering
    \includegraphics[width=\textwidth]{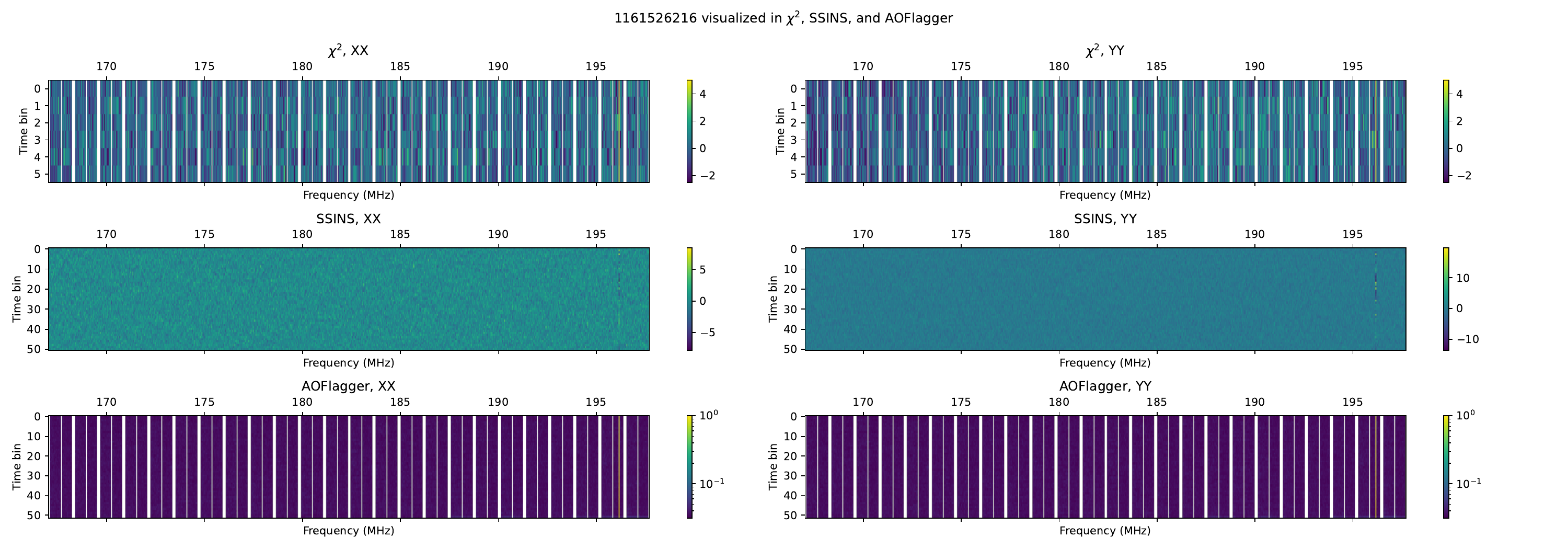}
    \caption{An example of narrow-band RFI at 196.175 MHz, detected by all three algorithms.}
    \label{fig:1161526216}
\end{figure*}

\begin{figure*}[ht]
    \centering
    \includegraphics[width=\textwidth]{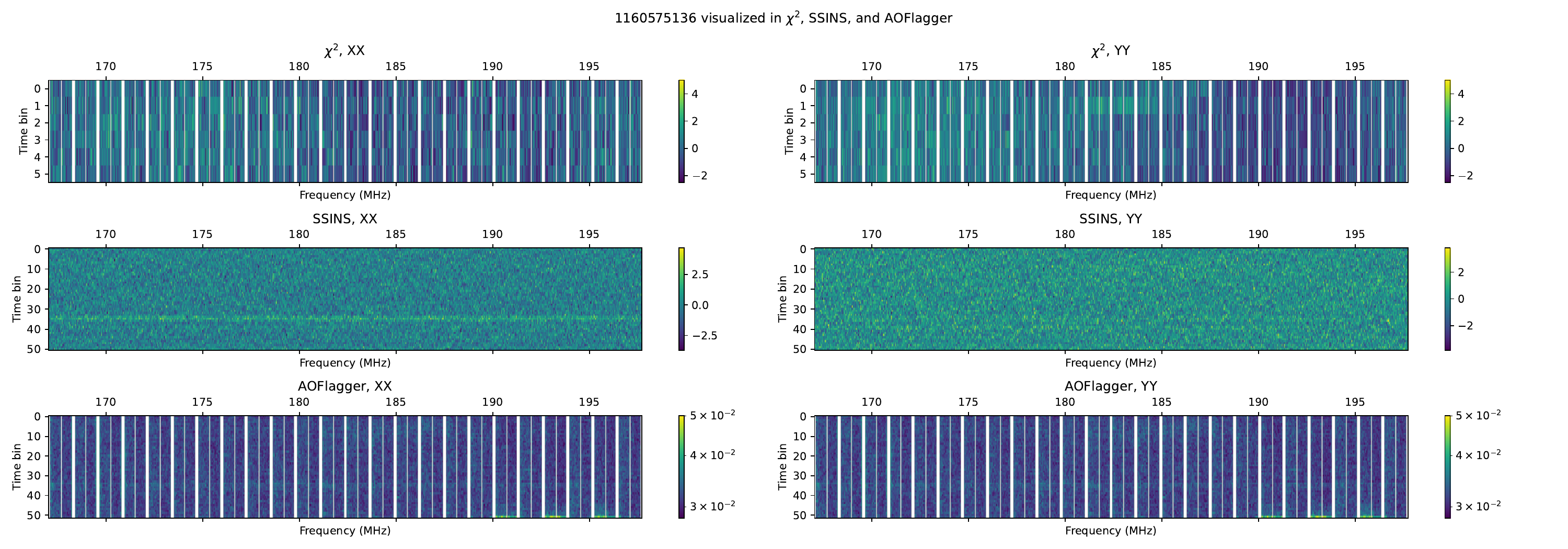}
    \caption{An example of a ``streak'' found in SSINS, but invisible to $\chi^2$.}
    \label{fig:1160575136}
\end{figure*}

\begin{figure*}[ht]
    \centering
    \includegraphics[width=\textwidth]{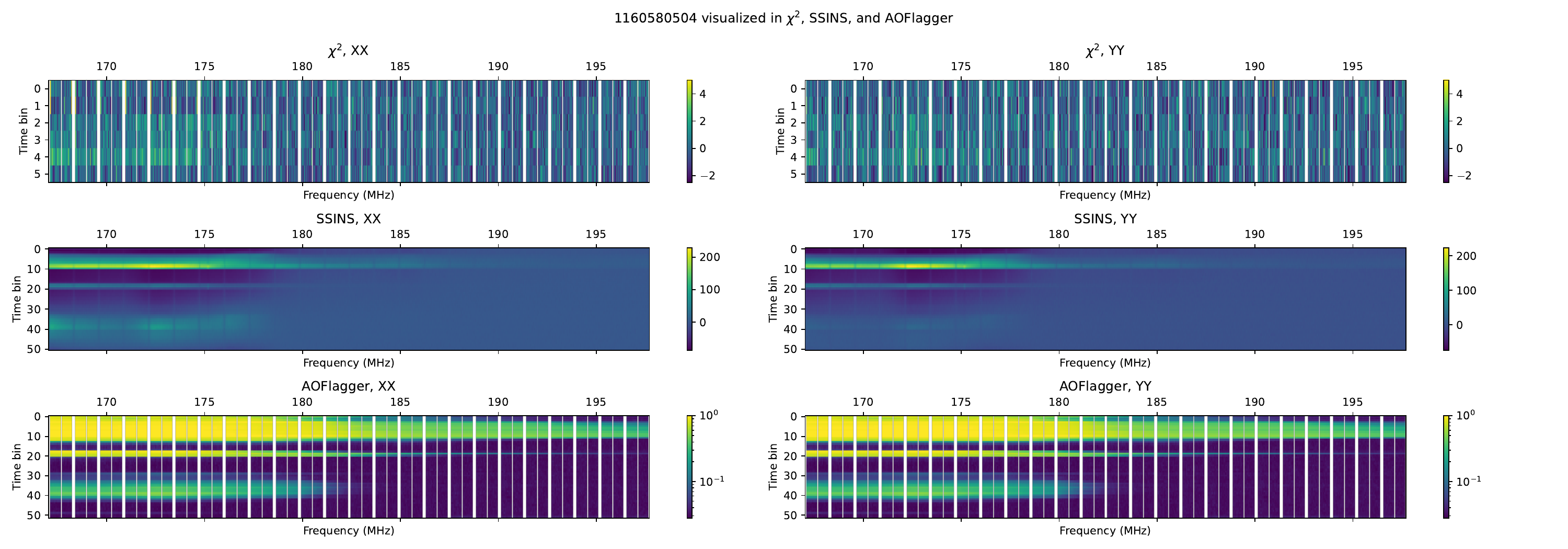}
    \caption{An unusual event, which might have been due to weather, is also nearly invisible to $\chi^2$.}
    \label{fig:1160580504}
\end{figure*}

Figures \ref{fig:chisq_only} through \ref{fig:1160580504} show time-frequency waterfall plots of $\chi^2/\mathrm{nDoF}$ modified z-scores (top row), SSINS's Incoherent Noise Spectrum waterfall plot (INS, middle row), and AOFlagger flags averaged over baseline as described in section \ref{aoflagger} (bottom row).  Each column shows one of the two linear polarizations measured by the MWA. 

\end{document}